\documentclass{article}
\usepackage{graphicx}
\usepackage{nopageno}

\input{tcilatex}

\begin{document}

\ \ \ \ \ \ \ \ \ \ \ \ \ {\LARGE Quantum correction with three codes }

\ \ \ \ \ \ \ \ \ \ \ \ \ \ \ \ \ \ {\LARGE \ \ \ \ }

\begin{center}
$^{{\small 1}}${\Large Aziz Mouzali, }$^{{\small 2}}${\Large Fatiha Merazka }

$^{{\small 1}}${\normalsize Faculty of Sciences, University of Blida, Alg%
\'{e}ria.}

$^{{\small 2}}Electrnic\&${\normalsize Computer engineering Faculty,
University of Science \& Technology Houari Boumediene, Algeria. fmerazka@usthb.dz\ \ \ \ \ \ \
\ \ \ \ \ \ \ \ \ \ \ \ \ \ \ }
\end{center}

\bigskip

\ \ \ \ \ \ \ \ \ \ \ \ \ \ \ \ \ \ \ \ \ \ \ \ \ \ \ \ \ \ \ \ \ \ \ \ \ \
\ \ \ \ {\Large Abstract\ \ \ \ \ \ \ \ \ \ \ \ \ \ }{\LARGE \ \ \ \ }

\begin{quote}
\ \ \ \ \ \ In this paper, we provide an implementation of five, seven and
nine-qubits error correcting codes on a classical computer using the quantum
simulator Feynman program.\ We also compare the three codes by computing the
fidelity when double errors occurs in a depolarizing channel. As triple
errors and more are considered very unlikely, it has negligible effect on
the next results.\ \ \

\bigskip


\bigskip

{\LARGE \ }
\end{quote}

\noindent {\Large 1 Introduction \ \ }

{\large \ \ \ }

The classical information processing is performed with physical supports,
which are governed by the \hspace{0cm}classical physics laws. The quantum
information processing uses particles, which obeys to quantum mechanics
(electrons, ions, photons...). An electron can carry information : according
to its spin, it represents the bit 0 or the bit 1. The quantum information
support called qubit is a physical system, which can be in two quantum
states representing the bits 0 or 1[ ]. The advantage of quantum information
processing is the possibility for the qubit to be in a two states
superposition, which is forbidden for the classical bit. \ However, any
environment qubit interaction will break this superposition at one of its
two possible states. Another essential tool is the entanglement between the
qubits, which is used in quantum computing. An n-qubits system entangled
state is not the n-tensor\ product\ of every qubit state, as it is the case
for a separable state. Unfortunately, the entanglement is very sensitive to
the environment noise. The superposition and entanglement loss is called
decoherence and introduces errors in computation, which requires quantum
error correcting code. In fact, while the classical computation error
probability is negligible, the error probability is very high in quantum
computation [ ]. The difference comes from the physical nature of the
information support : the electrons number in a capacitor\ can vary without
changing the bit, but if an electron spin changes then the bit value will
change. As the experimental protection against decoherence is very hard,
several quantum codes have been built to correct computation errors. A
quantum error correcting code aims to identify the qubits states alteration
caused by the noises and then to recover the correct states. A quantum code
is similar to a classical code, but ha$s$ one particularity: We cannot copy
the bit to protect it because it is impossible to clone the quantum
information. Also, we can not read out directly the information contained in
the useful qubit.\ In fact, we intricate the superposed qubit state carrying
the useful information with a number of additional qubits states, called
ancillas, to build the codedwords. If an errors affects the qubits, a
correcting procedure allows getting a system separable state, containing the
useful information {\normalsize [1]}. \bigskip

We propose in this paper, to simulate the five, seven and nine qubits code
This paper is organized as follows.In section 2, we take a look at some basics of quantum error correcting
codes, bit flip and phase flip correction and error measurement . In section
3, we describe the bit flip, phase flip and the Shor's nine qubits codes.
Sections 4, 5 and 6 are dedicated to the correction of double errors
respectively by the nine, seven and five qubits codes. Section 7 is devoted
to the definition of fidelity used in this work, then in Section 8, we
present and discuss the experimental results of this fidelity when double
errors occur in depolarizing channel and corrected by the five, seven and
nine qubits codes. Finally, we offer our conclusions in Section 9.

\bigskip

\noindent $\noindent $\noindent {\Large 2 Basics of Quantum Error Correction
\ }\bigskip

\noindent

Consider an n-qubits system in interaction with the environment.\ We
represent an error as unitary transformation U over all the
"n-qubits+environment" state. This error can be a bit flip X, a phase flip Z
and both of them Y=iXZ. We give below, the matricial representation of these
errors and their effect on a one superposed state $\alpha \left\vert
0\right\rangle +\beta \left\vert 1\right\rangle $. The Kets $\left\vert
0\right\rangle $ and $\left\vert 1\right\rangle $ correspondent, for
example, to respectively the spin +1/2 and -1/2 of an electron , or the
fundamental and exited states of an atom. The factors $\alpha $ and $\beta $
are in general complex number and give the probability $\left\vert \alpha
\right\vert ^{2}$ and\ $\ \left\vert \beta \right\vert ^{2}$ $(\left\vert
\alpha \right\vert ^{2}+\left\vert \beta \right\vert ^{2}=1)$ that the qubit
is respectively in the states $\left\vert 0\right\rangle $ and $\left\vert
1\right\rangle ${\large . }

\bigskip

\noindent If none error occurs, we use the identity matrix I:\

\bigskip

\noindent\ $\ \ I\left( \alpha \left\vert 0\right\rangle +\beta \left\vert
1\right\rangle \right) =\alpha \left\vert 0\right\rangle +\beta \left\vert
1\right\rangle $ \ \ \ or\noindent\ \ \ \ \noindent $I\left(
\begin{array}{c}
\alpha \\
\beta%
\end{array}%
\right) =\left(
\begin{array}{cc}
1 & 0 \\
0 & 1%
\end{array}%
\right) \left(
\begin{array}{c}
\alpha \\
\beta%
\end{array}%
\right) =\left(
\begin{array}{c}
\alpha \\
\beta%
\end{array}%
\right) $

\noindent

\bigskip

\noindent The Bit flip error is represented by the X Pauli matrix :$\bigskip
$

\noindent

\noindent\ $\ \ X\left( \alpha \left\vert 0\right\rangle +\beta \left\vert
1\right\rangle \right) =\beta \left\vert 0\right\rangle +\alpha \left\vert
1\right\rangle $ \ \ or$\bigskip $ \noindent $\ \ \ X\left(
\begin{array}{c}
\alpha \\
\beta%
\end{array}%
\right) =\left(
\begin{array}{cc}
0 & 1 \\
1 & 0%
\end{array}%
\right) \left(
\begin{array}{c}
\alpha \\
\beta%
\end{array}%
\right) =\left(
\begin{array}{c}
\beta \\
\alpha%
\end{array}%
\right) $

\bigskip \noindent

\noindent The phase flip error is represented by the Z \ Pauli matrix :$%
\bigskip $\noindent

\noindent\ \ \ \noindent $Z\left( \alpha \left\vert 0\right\rangle +\beta
\left\vert 1\right\rangle \right) =\alpha \left\vert 0\right\rangle -\beta
\left\vert 1\right\rangle $ or $Z\left(
\begin{array}{c}
\alpha \\
\beta%
\end{array}%
\right) =\left(
\begin{array}{cc}
1 & 0 \\
0 & -1%
\end{array}%
\right) \left(
\begin{array}{c}
\alpha \\
\beta%
\end{array}%
\right) =\left(
\begin{array}{c}
\alpha \\
-\beta%
\end{array}%
\right) $ \

\noindent

\bigskip \noindent The Bit and phase flip error is represented by the Y=-iXZ
\ Pauli matrix :$\bigskip $\noindent

\noindent$Y(\alpha \left\vert 0\right\rangle +\beta \left\vert
1\right\rangle )=i(-\beta $ $\left\vert 0\right\rangle +\alpha \left\vert
1\right\rangle )$ or\noindent \bigskip\ \noindent $Y\left(
\begin{array}{c}
\alpha \\
\beta%
\end{array}%
\right) =i\left(
\begin{array}{cc}
0 & -1 \\
1 & 0%
\end{array}%
\right) \left(
\begin{array}{c}
\alpha \\
\beta%
\end{array}%
\right) =i\left(
\begin{array}{c}
-\beta \\
\alpha%
\end{array}%
\right) $\bigskip

We also use in quantum error correction the Hadamard (H), controlled-not
(CNOT or CN) and Toffoli (T) gates. They act respectively on one, two and
three qubits. We give below their matricial representation and effect on
some states. \ \bigskip

\noindent $\noindent \noindent ${\Large Hadamard gate :} \ \ \ ~$H=\frac{1}{%
\sqrt{2}}\left(
\begin{array}{cc}
1 & 1 \\
1 & -1%
\end{array}%
\right) $\ \

\bigskip \noindent This gate transforms a simple state in a superposed state
and inversely:$\bigskip $ \

\noindent $H\left\vert 0\right\rangle =\frac{1}{\sqrt{2}}\left( \left\vert
0\right\rangle +\left\vert 1\right\rangle \right) $\bigskip\ $=\left\vert
+\right\rangle ;$ \ $H\left\vert 1\right\rangle =\frac{1}{\sqrt{2}}\left(
\left\vert 0\right\rangle -\left\vert 1\right\rangle \right) =\left\vert
-\right\rangle ;$ \

$\ $

\noindent $H\left\vert +\right\rangle =\left\vert 0\right\rangle ;$ \ $%
H\left\vert -\right\rangle =\left\vert 1\right\rangle \bigskip $ \ \ \ \

\noindent $\noindent $\noindent $\noindent \noindent ${\Large Controlled-not
gate (CNOT) : }$\ \ \ CN=\left(
\begin{array}{cccc}
1 & 0 & 0 & 0 \\
0 & 1 & 0 & 0 \\
0 & 0 & 0 & 1 \\
0 & 0 & 1 & 0%
\end{array}%
\right) $

\noindent

\noindent\ \ \ \ \ \ This gate acting on two qubits, flips the second one
(target) qubit only if the first one (controller) is equal to 1: \ $\bigskip
$

\noindent $CN\left\vert 00\right\rangle =\left\vert 00\right\rangle $ \ \ ;
\ $CN\left\vert 01\right\rangle =\left\vert 01\right\rangle $ \ ; \ $%
CN\left\vert 10\right\rangle =\left\vert 11\right\rangle $ \ \ ; \ \ $%
CN\left\vert 11\right\rangle =\left\vert 10\right\rangle $\bigskip\ $\ $

\noindent

\noindent $\noindent $\noindent $\noindent \noindent ${\Large Toffoli gate
(CCN) :}{\large \ } $T=\left(
\begin{array}{cccccccc}
1 & 0 & 0 & 0 & 0 & 0 & 0 & 0 \\
0 & 1 & 0 & 0 & 0 & 0 & 0 & 0 \\
0 & 0 & 1 & 0 & 0 & 0 & 0 & 0 \\
0 & 0 & 0 & 1 & 0 & 0 & 0 & 0 \\
0 & 0 & 0 & 0 & 1 & 0 & 0 & 0 \\
0 & 0 & 0 & 0 & 0 & 1 & 0 & 0 \\
0 & 0 & 0 & 0 & 0 & 0 & 0 & 1 \\
0 & 0 & 0 & 0 & 0 & 0 & 1 & 0%
\end{array}%
\right) $\bigskip

\noindent\ \ \ \ \ \ This gate acting on three qubits, flips the third one
(target) only if the two others are equal to 1: $\ $\bigskip

\noindent $\noindent \noindent \noindent \noindent T\left\vert
110\right\rangle =\left\vert 111\right\rangle $ \ \ \ \ $;$ \ \ \ \ $%
T\left\vert 111\right\rangle =\left\vert 110\right\rangle $\ \ \ \ \ \ \

\bigskip \noindent {\Large \ }\noindent $\noindent $\noindent {\Large \ }%
\newpage

\noindent \noindent \noindent {\Large 3 Quantum error correcting codes }

\bigskip

We present below some quantum error correcting codes (QECC) used to protect
useful information stored in a superposed qubit state $\left\vert \Psi
\right\rangle =\alpha \left\vert 0\right\rangle +\beta \left\vert
1\right\rangle .$

\bigskip

\bigskip

\noindent $\noindent $\noindent {\Large 3.1 Bit flip code}

\bigskip

To correct a bit flip, we use a redundant code by coding $\left\vert
0\right\rangle $ as $\left\vert 000\right\rangle $ and $\left\vert
1\right\rangle $ as $\left\vert 111\right\rangle .$The initial state is
coded as $\alpha \left\vert 000\right\rangle +\beta \left\vert
111\right\rangle $ and becomes after error on, for example, the third qubit $%
\alpha \left\vert 001\right\rangle +\beta \left\vert 110\right\rangle .$
W{}e have supposed that a bit flip occurs on all the superposition terms.\
To identify the affected qubit we add a fourth one initialized to $%
\left\vert \Psi _{4}\right\rangle =\left\vert 0\right\rangle $ which will
locate and record its position. The system state will be then $\left( \alpha
\left\vert 001\right\rangle +\beta \left\vert 110\right\rangle \right)
\left\vert 0\right\rangle $. We determine by calculus the error position and
obtain the state $\left( \alpha \left\vert 001\right\rangle +\beta
\left\vert 110\right\rangle \right) \left\vert \Psi _{4}^{3}\right\rangle $
where the fourth qubit state $\left\vert \Psi _{4}^{3}\right\rangle $
indicates that the third qubit is affected by an error. After correcting the
designed qubit, the global state becomes $\left( \alpha \left\vert
000\right\rangle +\beta \left\vert 111\right\rangle \right) \left\vert \Psi
_{4}^{3}\right\rangle .$ Finally, we suppress the redundancy and return to
the initial state $\alpha \left\vert 0\right\rangle +\beta \left\vert
1\right\rangle $ {\large .}\ This procedure is also valid when the error
occurs on superposition. Consider a noise which produces the state $\left(
\alpha \left\vert 100\right\rangle +\beta \left\vert 011\right\rangle
+\alpha \left\vert 001\right\rangle +\beta \left\vert 110\right\rangle
\right) \left\vert 0\right\rangle .$ Adding the error register and locating
the affected qubit gives $\frac{1}{\sqrt{2}}\left[ \left( \alpha \left\vert
100\right\rangle +\beta \left\vert 011\right\rangle \right) \left\vert
1\right\rangle +\left( \alpha \left\vert 001\right\rangle +\beta \left\vert
110\right\rangle \right) \left\vert \Psi _{4}^{13}\right\rangle \right] $
where the state $\left\vert \Psi _{4}^{13}\right\rangle $ indicates that the
first and the third qubits are affected by errors. Measuring the\ error
register gives $\left( \alpha \left\vert 100\right\rangle +\beta \left\vert
011\right\rangle \right) \left\vert \Psi _{4}^{1}\right\rangle $ or $\left(
\alpha \left\vert 001\right\rangle +\beta \left\vert 110\right\rangle
\right) \left\vert \Psi _{4}^{3}\right\rangle .$ Finally, we flip the qubits
designed by the error register to obtain the state $\alpha \left\vert
000\right\rangle +\beta \left\vert 111\right\rangle .$ Then, we suppress the
redundancy to return to the original state $\alpha \left\vert 0\right\rangle
+\beta \left\vert 1\right\rangle $ {\large [2-4].}\bigskip

Figure 1 shows the Bit flip code{\Large \ }circuit. The encoding and
decoding \ circuits are located in the time intervals $\left[ t_{0\text{ }%
},t_{1\text{ }}\right] $ and $\left[ t_{2\text{ }},t_{3\ \text{ }}\right] $,
respectively. The useful information is stored on the first qubit state $%
\left\vert \Psi \right\rangle =\alpha \left\vert 0\right\rangle +\beta
\left\vert 1\right\rangle $ . The second and third qubits are added\ to
recover this state if a bit flip occurs in $\left[ t_{1\text{ }},t_{2\ \text{
}}\right] ${\large [2-4].}$\bigskip $

$\ \ \ \ \ \ \ \ \ \ \
$

$\ \ \ \ \ \ \ \ \ \ $\bigskip

\noindent\ \ \ \ \noindent At time t$_{\text{0}}$, the initial separable
state system is :\noindent $\bigskip $

\noindent\ $\ \ \ \ \ \ \ \ \ \ \ \ \ \ \ \ \ \ \ \ \ \ \ \left\vert \Psi
_{0}\right\rangle =\left\vert \Psi \right\rangle \left\vert 0\right\rangle
\left\vert 0\right\rangle =\alpha \left\vert 000\right\rangle +\beta
\left\vert 100\right\rangle $\bigskip \noindent\ \ \ \ \ \ \ \ \ \ \ \ \ \ \
\ \ \ \ \ \ \ \ \ \ \ \ \ \ \ \ \ \ \ \ \ \ $\mathbf{(1)}$

\noindent\ \ \ \ \ \ After applying the coding and decoding circuit we
obtain at time t$_{\text{3}}$ , the disintricated state:\bigskip\ \

\noindent\ $\ \ \ \ \ \ \ \ \ \ \ \ \ \ \ \ \ \ \ \ \left\vert \Psi
_{3}\right\rangle =\alpha \left\vert 011\right\rangle +\beta \left\vert
111\right\rangle =\left( \alpha \left\vert 0\right\rangle +\beta \left\vert
1\right\rangle \right) \left\vert 11\right\rangle $ \ \ $\ \ \ \ \ \ \ \ \ \
\ \ \ \ \ \ \ \ \ \ \ \ \ \ \ \mathbf{(2)}$

\bigskip

We have then, recovered the first qubit state, which contains the useful
information. The bit flip in the other two qubits indicates, that a bit flip
has affected the first one. If a bit flip occurs on the second and third
qubits, we obtain the states, respectively as:\bigskip

\noindent $\noindent \noindent $\noindent $\left\vert \Psi _{2}\right\rangle
=\alpha \left\vert 010\right\rangle +\beta \left\vert 101\right\rangle $%
{\tiny ;\ }$\left\vert \Psi _{2}^{^{\prime }}\right\rangle =\alpha
\left\vert 010\right\rangle +\beta \left\vert 110\right\rangle ${\tiny ;\ \ }%
$\left\vert \Psi _{3}\right\rangle =\alpha \left\vert 0\right\rangle +\beta
\left\vert 1\right\rangle \left\vert 10\right\rangle ${\tiny \bigskip\ \ \ \
\ \ \ \ \ \ \ \ \ \ \ \ \ \ \ \ \ }$\mathbf{(3)}$

\noindent $\noindent \noindent \left\vert \noindent \Psi _{2}\right\rangle
=\alpha \left\vert 001\right\rangle +\beta \left\vert 110\right\rangle $%
{\tiny ; }$\left\vert \Psi _{2}^{^{\prime }}\right\rangle =\alpha \left\vert
001\right\rangle +\beta \left\vert 101\right\rangle ${\tiny ;\ \ }$%
\left\vert \Psi _{3}\right\rangle =\alpha \left\vert 0\right\rangle +\beta
\left\vert 1\right\rangle \left\vert 01\right\rangle $\bigskip\ \ \ \ \ $\ \
\ \ \ \mathbf{(4)}$

Appendix B shows the bit flip code simulation made with the Feynman program.
The procedure called "Bitflip" takes as an input the bit flip affected\
qubit\ number n=1, 2 or 3. The input n=0 corresponds to the case where no
errors occur.

\emph{\ \ \ \ }

\bigskip

\emph{\ \ \ \ \ \ \ \ \ \ \ }\bigskip

\noindent {\Large 3.2 Phase flip code }\bigskip

We want to protect the qubit state\ $\alpha \left\vert 0\right\rangle +\beta
\left\vert 1\right\rangle $ from a phase flip error. The idea is to
transform a phase flip, which is uncorrectable, into a bit flip, which is
correctable. This state is then coded as $\alpha \left\vert +++\right\rangle
+\beta \left\vert ---\right\rangle $ where $\left\vert +\right\rangle $ and $%
\left\vert -\right\rangle $ are obtained by the Hadamard gate:\ $H\left\vert
0\right\rangle =\frac{1}{\sqrt{2}}\left( \left\vert 0\right\rangle
+\left\vert 1\right\rangle \right) =\left\vert +\right\rangle $\ and\ $%
H\left\vert 1\right\rangle =\frac{1}{\sqrt{2}}\left( \left\vert
0\right\rangle -\left\vert 1\right\rangle \right) =\left\vert -\right\rangle
.$ A phase flip on the first qubit gives the state $\alpha \left\vert
-++\right\rangle +\beta \left\vert +--\right\rangle .$ The correction begins
by a three transformation $H^{\otimes 3}$ which allows to return to the
basic state $H^{\otimes 3}\alpha \left\vert -++\right\rangle +\beta
\left\vert +--\right\rangle =\alpha \left\vert 100\right\rangle +\beta
\left\vert 011\right\rangle .$ Then we correct the affected qubit to restore
the state $\alpha \left\vert 000\right\rangle +\beta \left\vert
111\right\rangle .$ Finally, we apply $H^{\otimes 3}$ to return at the coded
state $\alpha \left\vert +++\right\rangle +\beta \left\vert ---\right\rangle
$ {\large [2-4]. }\ \newline

Figure 2 shows the phase flip code circuit. The useful information is stored
on the first qubit state $\left\vert \Psi \right\rangle =\alpha \left\vert
0\right\rangle +\beta \left\vert 1\right\rangle $.\ The encoding and
decoding \ circuits are located in the time intervals $\left[ t_{0\text{ }%
},t_{1\ \text{ }}\right] $and $\left[ t_{2\text{ }},t_{3\ \text{ }}\right] $%
, respectively. This code uses three qubits to correct a phase flip, that
affect one of them between t$_{\text{1}}\ $and t$_{\text{2 }}$ {\large [2-4].%
}

\bigskip


\bigskip

\noindent\ \ \ \ \ \ After applying the coding and decoding circuit we
obtain at time t$_{\text{3}}$ , the disintricated state :\bigskip\ \ \ $\ $

\noindent\ \ \ \ \ \ \ \ \ \ \ \ \ \ \ \ \ \ \ \ \ \ \ \ \ \ \ \ \ \ \ \ \ \
\ \ \ \ \ \ \noindent $\noindent \left\vert \noindent \Psi _{3}\right\rangle
=\frac{1}{\sqrt{2}}\left[ \alpha \ \left\vert 0\right\rangle +\beta \
\left\vert 1\right\rangle \right] \left\vert 11\right\rangle =\left\vert
\Psi \right\rangle \left\vert 11\right\rangle {\tiny \bigskip }$ \ \ \ \ \ \
\ \ \ \ \ \ \ \ \ \ \ \ \ $\mathbf{(5)}$\bigskip

We have then, recovered the first qubit state, which contains the useful
information. The bit flip in the other two qubits indicates, that a phase
flip has affected the first one.\bigskip \bigskip\

Appendix B shows the phase flip code simulation made with the Feynman
program. The procedure called "Bitflip" takes as an input the phase flip
affected\ qubit\ number n=1, 2 or 3. The input n=0 corresponds to the case
where no errors occur. We give, below, the outputs for the fourth n values.
The running time is about two seconds.\ \ \ \ \ \ \ \ \ \ \ \ \ \ \ \ \ \ \
\ \ \ \ \ \ \ \ \ \ \ \ \ \ \ \ \ \ \ \ \ \ \ \ \ \ \ \ \ \ \ \ \ \ \ \ \ \
\ \ \ \ \ \ \ \ \ \ \ \ \ \ \ \ \ \ \ \ \ \ \ \ \ \ \ \ \ \ \ \ \ \ \ \ \ \
\ \ \ \ \ \ \ \ \ \ \ \ \ \ \ \ \ \ \ \ \ \ \ \ \ \ \ \ \ \ \ \ \ \ \ \ \ \
\ \ \ \ \ \ \ \ \ \ \ \ \ \ \ \ \ \ \ \ \ \ \ \ \ \ \ \ \ \ \ \ \ \

\noindent\ \emph{\bigskip }

\noindent {\Large 3.3 Shor code}\bigskip

The Shor code uses nine qubits to correct a combined phase and bit flip
error on one qubit. We describe below the encoding, correction and decoding
procedures of this code.\noindent

\bigskip

\bigskip

\noindent{\Large 3.3.1 Encoding }\bigskip

The initial state \ $\alpha \left\vert 0\right\rangle +\beta \left\vert
1\right\rangle $ is coded $\alpha \left\vert +++\right\rangle +\beta
\left\vert ---\right\rangle =\alpha \left( \frac{1}{\sqrt{2}}\left(
\left\vert 0\right\rangle +\left\vert 1\right\rangle \right) \right)
^{\otimes 3}+\beta \left( \frac{1}{\sqrt{2}}\left( \left\vert 0\right\rangle
-\left\vert 1\right\rangle \right) \right) ^{\otimes 3}$ so it could be
protected against phase flip. Then we protect against bit flip by coding
each $\left\vert 0\right\rangle $ as $\left\vert 000\right\rangle $ and each
$\left\vert 1\right\rangle $ as $\left\vert 111\right\rangle $ and obtain :

$\alpha \left( \frac{1}{\sqrt{2}}\left( \left\vert 000\right\rangle
+\left\vert 111\right\rangle \right) \right) ^{\otimes 3}+\ \beta \left(
\frac{1}{\sqrt{2}}\left( \left\vert 000\right\rangle -\left\vert
111\right\rangle \right) \right) ^{\otimes 3\text{ }}.$ The Shor algorithm,
which will be described below, corrects first the bit flip, then the phase
flip {\large [2][3][4].}

\bigskip

Figure 3 shows the circuit of this code\ {\large [4]. }This code uses nine
qubits to correct a single X, Y or Z error on only one of them. It is the
concatenation of the two codes described above: the bit and the phase flip
codes {\large [4]}.\textbf{%
}{\large \ }

\bigskip

The useful information is stored on the the first qubit state $\left\vert
\Psi \right\rangle =\alpha \left\vert 0\right\rangle +\beta \left\vert
1\right\rangle $.\textbf{\ }The others ones, are added\ to recover this
state if an error occurs between the instants t$_{\text{1}}$ and\ t$_{\text{2%
}}.\ $We can remove or use them again , after recovering their initial state
$\left\vert 0\right\rangle $. Boxes C$_{\text{B}}$\ and D$_{\text{B}}$\ are
called coding and decoding boxes, respectively. They correspond to the
circuits shown on Figure 4. Symbol D indicates the error detection measured
on the six additional qubits :%

{\Huge \ }

\noindent

After applying the coding circuit, we obtain at time t$_{\text{1}}$ the
coded state:\bigskip\ \noindent

$\ \ \ \ \ \ \ \ \ \ \ \ \ \ \left\vert \Psi _{1}\right\rangle =\frac{1}{%
2^{3/2}}\left[ \alpha (\left\vert 000\right\rangle +\left\vert
111\right\rangle )^{\otimes 3}+\beta (\left\vert 000\right\rangle
-\left\vert 111\right\rangle )^{\otimes 3}\right] $ \ \ \ \ \ \ \ \ \ \ \ \
\ \ \ \ \ \ \ \ $\mathbf{(6)}$\bigskip \noindent \noindent\

\noindent or $\ \ \ \ \ \ \ \ \ \ \ \ \ \ \ \ \ \ \ \ \ \ \ \ \ \ \ \ \ \ \
\ \ \ \left\vert \Psi _{1}\right\rangle =\left[ \alpha \left\vert
0_{L}\right\rangle +\beta \left\vert 1_{L}\right\rangle \right] $\bigskip\ \
\ \ \ $\ \ \ \ \ \ \ \ \ \ \ \ \ \ \ \ \ \ \ \ \ \ \ \ \ \ \ \ \ \ \ \ \ \ \
\ \ \ \ \ \ \ \ \ \ \ \ \mathbf{(7)}$

The label 'L' indicates the logical qubit state, which is different from the
physical qubit state $\left\vert \Psi \right\rangle =\alpha \left\vert
0\right\rangle +\beta \left\vert 1\right\rangle $. The codewords of this
nine qubits code are:$\bigskip $

\ \ \ \ $\ \ \left\vert 0_{L}\right\rangle =\frac{1}{2^{3/2}}(\left\vert
000\right\rangle +\left\vert 111\right\rangle )^{\otimes 3}\ \ \ \ $and \ \
\ \ $\left\vert 1_{L}\right\rangle =\frac{1}{2^{3/2}}(\left\vert
000\right\rangle -\left\vert 111\right\rangle )^{\otimes 3}$ \ \ $\ \ \ \ \
\ \mathbf{(8)}$\bigskip \bigskip

\noindent {\Large 3.3.2 \ X error on the useful qubit}\noindent

\bigskip

If an a bit flip X occurs between t$_{1}$ and t$_{2}$ the affected
decodedstate obtained at time t$_{3}$ is :\bigskip

\noindent\ $\ \ \ \ \ \ \ \ \ \noindent \noindent \noindent \noindent
\left\vert \Psi _{3}\right\rangle =\left( \alpha \left\vert 000\right\rangle
+\beta \left\vert 100\right\rangle \right) \left\vert 11\right\rangle
(\left\vert 00\right\rangle )^{\otimes 2}=\left\vert \Psi \right\rangle
\left\vert 00\right\rangle \left\vert 11\right\rangle {\large (}\left\vert
00\right\rangle {\large )}^{\otimes 2}$ \ \ \ \ \ \ \ \ \ \ \ \ \ \ $\mathbf{%
(9)}$\bigskip

This final state is separable and the first qubit initial state, carrying
the useful information has been recovered. The measured modification on the
forth\ and fifth states qubits lets to know that\ an\ error\ occurred. The
additional qubits can now, be removed or used again, after putting them on
the $\left\vert 0\right\rangle $\ state.\noindent

\bigskip

\noindent {\Large 3.3.3\ \ Z error on the useful qubit\bigskip }

\bigskip\ If an a bit flip X occurs between t$_{1}$ and t$_{2}$ the affected
decodedstate obtained at time t$_{3}$ is :\bigskip

\bigskip\ \ \ \ \ \ \ \ \ \ \ \ \ \ \ \ \ \ \ \ \ \ \ \ \ \ \ \ \ \ \ \ \ \
\noindent $\noindent \noindent $\noindent $\left\vert \Psi _{3}\right\rangle
{\large =}\left\vert \Psi \right\rangle \left\vert 11\right\rangle {\large (}%
\left\vert 00\right\rangle {\large )}^{\otimes 3}$ \ \ \ \ \ \ \ \ \ \ \ \ \
\ \ \ \ \ \ \ \ \ \ \ \ \ $\ \ \ \ \ \ \ \ \ \ \ \ \ \ \mathbf{(10)}$\bigskip

The first qubit state is then recovered. The modification on the second and
third qubits states, indicates an error occurrence.\bigskip

\bigskip

\noindent {\Large 3.3.4\ Y error on the useful qubit\ \ }

{\Large \bigskip }

\noindent\ \ \ \ \ \ If an a bit flip X occurs between t$_{1}$ and t$_{2}$
the affected decodedstate obtained at time t$_{3}$ is :\bigskip

$\ \ \ \ \ \ \ \ \ \ \ \ \ \ \ \ \ \ \ \ \ \ \ \ \ \ \ \ \ \ \ \ \left\vert
\Psi _{3}\right\rangle =i\left\vert \Psi \right\rangle \left\vert
1111\right\rangle (\left\vert 00\right\rangle )^{\otimes 2}$\bigskip\ \ \ \
\ \ \ \ $\ \ \ \ \ \ \ \ \ \ \ \ \ \ \ \ \ \ \ \ \ \ \ \ \ \ \ \mathbf{(11)}$

The first qubit state is then recovered and the measured modification on the
second, third, forth and fifth qubit states indicates that an error has
occurred. We remark, that the states $\left\vert \Psi \right\rangle $ and $%
i\left\vert \Psi \right\rangle $ are physically the same, so we do not need
to suppress the complex number i.

\bigskip

\noindent {\Large 3.3.5\ Arbitrary error on the useful qubit \ }\bigskip

Suppose now that an arbitrary error E=c$_{\text{o}}$I+c$_{\text{x}}$X\ +c$_{%
\text{y}}$Y+c$_{\text{z}}$Z\ \ occurs on the first qubit so that $\left\vert
\Psi _{2}\right\rangle =E\left\vert \Psi _{1}\right\rangle $.\ \ The
coefficients c$_{\text{x}}$, c$_{\text{y}}$ and c$_{\text{z}}$ give
respectively, the probability $\left\vert c_{x}\right\vert ^{2},$ \ $%
\left\vert c_{y}\right\vert ^{2}$ and $\left\vert c_{z}\right\vert ^{2}$
that an X, Y or Z error affects the qubit. The same calculus allows to
obtain the state $\left\vert \Psi _{3}\right\rangle $ as a linear
combination of the correct state $\left\vert \Psi _{3}\right\rangle
=\left\vert \Psi \right\rangle (\left\vert 00\right\rangle )^{\otimes 4}$
and the three states $\left\vert \Psi _{3}\right\rangle $ calculated for
respectively, the simple X, Y or Z errors :\bigskip \noindent $\ $

\noindent\ $\ \ \ \ \ \ \ \ \noindent \noindent \noindent \left\vert \Psi
_{3}\right\rangle =\left\vert \Psi \right\rangle \left[ c_{o}\left\vert
0000\right\rangle +\ c_{x}\left\vert 0011\right\rangle +ic_{y}\left\vert
1111\right\rangle +c_{z}\left\vert 1100\right\rangle \right] (\left\vert
00\right\rangle )^{\otimes 2}$ \ \ \ \ \ \ \ \ \ \ $\mathbf{(12)}$\newline

{\tiny \bigskip }

\noindent {\Large 3.3.6 \ Detection and correction error before decoding
\bigskip }

To detect an error in the state $\left\vert \Psi _{2}\right\rangle ,$ we
need to locate the affected qubit state. We then correct it by making
detection measure before the decoding procedure. We use for detection the
stabilizer group $G=\left\{ g_{i},\text{ }i=1..8\right\} $ of the Shor code
which is:

\noindent $\left\{
Z_{1}Z_{2},Z_{2}Z_{3},Z_{4}Z_{5},Z_{5}Z_{6},Z_{7}Z_{8},Z_{8}Z_{9},X_{1}X_{2}X_{3}X_{4}X_{5}X_{6},X_{4}X_{5}X_{6}X_{7}X_{8}X_{9}\right\}
$. \ \

\noindent Every generator $g_{i}$ verify the relation $g_{i}\left\vert \Psi
_{1}\right\rangle =\left\vert \Psi _{1}\right\rangle $ with $\left\vert \Psi
_{1}\right\rangle =\left[ \alpha \left\vert 0_{L}\right\rangle +\beta
\left\vert 1_{L}\right\rangle \right] $\textbf{.}\bigskip

We begin by applying all the six first generators on $\left\vert \Psi
_{2}\right\rangle $ to verify that no bit flip occurred. We use for this,
the fact that if no error occurs ($\left\vert \Psi _{2}\right\rangle
=\left\vert \Psi _{1}\right\rangle )$ or if only a phase flip occurs in $%
\left\vert \Psi _{2}\right\rangle $ all the terms in the superposition are $%
\pm \left\vert 000\right\rangle $ or $\pm \left\vert 111\right\rangle ,$ so
that we obtain:\bigskip

\noindent ${\small Z}_{1}{\small Z}_{2}\left\vert \Psi _{2}\right\rangle
{\small =}$\noindent ${\small Z}_{2}{\small Z}_{3}\left\vert \Psi
_{2}\right\rangle {\small =}$\noindent ${\small Z}_{4}{\small Z}%
_{5}\left\vert \Psi _{2}\right\rangle {\small =}$\noindent ${\small Z}_{5}%
{\small Z}_{6}\left\vert \Psi _{2}\right\rangle {\small =}$\noindent $%
{\small Z}_{7}{\small Z}_{8}\left\vert \Psi _{2}\right\rangle {\small =}$%
\noindent ${\small Z}_{8}{\small Z}_{9}\left\vert \Psi _{2}\right\rangle
{\small =}\left\vert \Psi _{2}\right\rangle $ \ \ \ \ \ \ $\mathbf{(13)}$%
\bigskip \bigskip

\noindent For example, if an X error occurs on the first qubit, we obtain
:\bigskip

\noindent $\left\vert \Psi _{2}\right\rangle {\large =}\frac{1}{2^{3/2}}%
[\alpha (\left\vert 100\right\rangle $\noindent $+\left\vert
011\right\rangle )(\left\vert 000\right\rangle $\noindent $+\left\vert
111\right\rangle )^{\otimes 2}$\noindent $+\beta (\left\vert
100\right\rangle $\noindent $-\left\vert 011\right\rangle )(\left\vert
000\right\rangle $\noindent $-\left\vert 111\right\rangle )^{\otimes
2}]\bigskip $ \ \ \ \ \ \ $\mathbf{(14)}$\newline

\noindent\ $\ \ \ \ \ \ \ \ \ \ \ \ \ \ \ \ \ Z_{1}Z_{2}\left\vert \Psi
_{2}\right\rangle ${\tiny \noindent }$=$\noindent $-\left\vert \Psi
_{2}\right\rangle \ \ \ \ \ \ \ \ \ \ \ $and$\ \ \ \ \ Z_{2}Z_{3}\left\vert
\Psi _{2}\right\rangle \noindent \ =\noindent \ \left\vert \Psi
_{2}\right\rangle \bigskip \ \ \ \ \ \ \ \ \ \ \ \ \ \ \ \ \ \ \ \ \ \ \ \ \
\ \ \ \ \ \ \ \mathbf{(15)}$\newline
$\ \ \ \ \ \ \ \ \ \ \ \ \ \ \ \ \ \ \ \ \ \ \ \ \ \ $\ \ \ \ \ \ \ \ \ \ \
\ \ \ \ \ \ \ \ \ \ \ \ \ \ \ \ \ \ \ \ \ \ \ \ \ \ \ \ \ $\mathbf{\ }$

The results are depicted on the table\ 1, where the first column gives the
bit flip affected qubit number. The values +1 and --1 are the $Z_{i}Z_{i+1}$
operators eigenvalues when applied on the $\left\vert \Psi _{2}\right\rangle
$ state. The last column gives the correction recovering the original state $%
\left\vert \Psi _{1}\right\rangle .$

\bigskip\ \
\begin{tabular}{|l|l|l|l|l|l|l|l|}
\hline
\textbf{Qubit} & $\mathbf{Z}_{1}\mathbf{Z}_{2}$ & $\mathbf{Z}_{2}\mathbf{Z}%
_{3}$ & $\mathbf{Z}_{4}\mathbf{Z}_{5}$ & $\mathbf{Z}_{5}\mathbf{Z}_{6}$ & $%
\mathbf{Z}_{7}\mathbf{Z}_{8}$ & $\mathbf{Z}_{8}\mathbf{Z}_{9}$ & \textbf{%
Correction} \\ \hline
\textbf{\ \ 1} & \textbf{\ --1} & \textbf{+1} & \textbf{+1} & \textbf{+1} &
\textbf{+1} & \textbf{+1} & $\mathbf{\ \ \ \ \ \ X}_{1}$ \\ \hline
\textbf{\ \ 2} & \textbf{\ --1} & \ \textbf{--1} & \textbf{+1} & \textbf{+1}
& \textbf{+1} & \textbf{+1} & $\mathbf{\ \ \ \ \ \ X}_{2}$ \\ \hline
\textbf{\ \ 3} & \textbf{+1} & \ \textbf{--1} & \textbf{+1} & \textbf{+1} &
\textbf{+1} & \textbf{+1} & $\mathbf{\ \ \ \ \ \ X}_{3}$ \\ \hline
\textbf{\ \ 4} & \textbf{+1} & \textbf{+1} & \ \textbf{--1} & \textbf{+1} &
\textbf{+1} & \textbf{+1} & $\mathbf{\ \ \ \ \ \ X}_{4}$ \\ \hline
\textbf{\ \ 5} & \textbf{+1} & \textbf{+1} & \ \textbf{--1} & \ \textbf{--1}
& \textbf{+1} & \textbf{+1} & $\mathbf{\ \ \ \ \ \ X}_{5}$ \\ \hline
\textbf{\ \ 6} & \textbf{+1} & \textbf{+1} & \textbf{+1} & \ \textbf{--1} &
\textbf{+1} & \textbf{+1} & $\mathbf{\ \ \ \ \ \ X}_{6}$ \\ \hline
\textbf{\ \ 7} & \textbf{+1} & \textbf{+1} & \textbf{+1} & \textbf{+1} & \
\textbf{--1} & \textbf{+1} & $\mathbf{\ \ \ \ \ \ X}_{7}$ \\ \hline
\textbf{\ \ 8} & \textbf{+1} & \textbf{+1} & \textbf{+1} & \textbf{+1} &
\textbf{--1} & \ \textbf{--1} & $\mathbf{\ \ \ \ \ \ X}_{8}$ \\ \hline
\textbf{\ \ 9} & \textbf{+1} & \textbf{+1} & \textbf{+1} & \textbf{+1} &
\textbf{+1} & \ \textbf{--1} & $\mathbf{\ \ \ \ \ \ X}_{9}$ \\ \hline
\end{tabular}

\bigskip\ \ \ \textbf{Table 1: Action of the generators on a bit flip
affected qubit. }

\bigskip

The obtained eigenvalues allow to identify the affected qubit 'i' and then
to correct it by applying the X$_{\text{i}}$ operator. In the second step,
we detect the phase flip by applying on $\left\vert \Psi _{2}\right\rangle $
the two generators $g_{7}=$ $X_{1}X_{2}X_{3}X_{4}X_{5}X_{6}\ \ \ \ $and%
\textbf{\ \ \ } $g_{8}=X_{4}X_{5}X_{6}X_{7}X_{8}X_{9\text{ \ }}$.\bigskip

\noindent For example, If a phase flip occurs on the first qubit we
obtain:\bigskip

\begin{center}
\noindent $\left\vert \Psi _{2}\right\rangle $\noindent $=$\noindent $\frac{1%
}{2^{3/2}}[\alpha (\left\vert 000\right\rangle $\noindent $-\left\vert
111\right\rangle )(\left\vert 000\right\rangle $\noindent $+\left\vert
111\right\rangle )^{\otimes 2}$\noindent $+\beta (\left\vert
000\right\rangle $\noindent $+\left\vert 111\right\rangle )(\left\vert
000\right\rangle $\noindent $-\left\vert 111\right\rangle )^{\otimes 2}]$%
\noindent \bigskip\ \ $\ \mathbf{(17)}$\\[0pt]
$\ \ \ \ \ \ \ \ \ \ \ \ \ \ \ \ \ \ \ \ \ \ \ g_{7}\left\vert \Psi
_{2}\right\rangle $\noindent $=$\noindent $\ $\noindent $-\left\vert \Psi
_{2}\right\rangle $ \ \ \ \ \ \ \ and \ \ \ \ \ \ $g_{8}\left\vert \Psi
_{2}\right\rangle $\noindent $=$ \noindent $\left\vert \Psi
_{2}\right\rangle $\ \ \ \ \ \ \ \ \ \ \ \ \ \ \ $\ \ \ \ \ \ \mathbf{(18)}$%
\noindent\ $\ \ \ \ \ \ \ \ \ \ \ \ \ \ \ \ \ \ \ \ \ \ \ \ \ \ \ \ \ \ \ \
\ \ \ \ \ \ \ \ \ \ \ \ \ \ \ \ \ \ \ \ \ \ \ \ \ \ \ \ \ \ \ \ \ \ $ $\ \ \
$
\end{center}

$\ \ \ \ \ \ \ \ \ \ \ \ \ \ \ \ \ \ \ \ \ \ $\ \ \ $\ \ \ \ \ \ \ \ \ \ \ \
\ \ \ \ \ \ $\ \ \ \ \ \ \ \ \ \ \ \ \ \ \ \ \ \ \ \ \ \ \ \ \ \

\noindent\ \ \ \ \ The results are summarized on the table\ 2, where the
first column gives the phase flip affected qubit number. The values +1 and
--1 are the eigenvalues of the operators $g_{7}$ and $g_{8}$ when applied on
the $\left\vert \Psi _{2}\right\rangle $ state. The last column gives the
correction, which recovers the original state $\left\vert \Psi
_{1}\right\rangle .$We note that this correction procedure works only if a
single error occurs and does not work in the case of an arbitrary error.

\bigskip\ \ \ \ \ \ \ \ \ \ \ \ \ \ \ \ \ \ \ \ \ \
\begin{tabular}{|l|l|l|l|}
\hline
\textbf{Qubit} & $\mathbf{\ \ \ \ g}_{7}$ & $\mathbf{\ \ \ g}_{8}$ & \textbf{%
Correction} \\ \hline
\textbf{\ \ \ 1} & \textbf{\ \ --1} & \textbf{\ \ +1 \ \ } & $\mathbf{\ \ \ Z%
}_{1}\mathbf{Z}_{2}\mathbf{Z}_{3}$ \\ \hline
\textbf{\ \ \ 2} & \textbf{\ \ --1} & \textbf{\ \ +1} & \ \textbf{\ }$\
\mathbf{Z}_{1}\mathbf{Z}_{2}\mathbf{Z}_{3}$ \\ \hline
\textbf{\ \ \ 3} & \textbf{\ \ --1} & \textbf{\ \ +1} & $\mathbf{\ \ \ Z}_{1}%
\mathbf{Z}_{2}\mathbf{Z}_{3}$ \\ \hline
\textbf{\ \ \ 4} & \textbf{\ \ --1} & \textbf{\ \ \ --1} & $\mathbf{\ \ \ Z}%
_{4}\mathbf{Z}_{5}\mathbf{Z}_{6}$ \\ \hline
\textbf{\ \ \ 5} & \textbf{\ \ --1} & \textbf{\ \ \ --1} & $\mathbf{\ \ \ Z}%
_{4}\mathbf{Z}_{5}\mathbf{Z}_{6}$ \\ \hline
\textbf{\ \ \ 6} & \textbf{\ \ --1} & \textbf{\ \ \ --1} & $\mathbf{\ \ \ Z}%
_{4}\mathbf{Z}_{5}\mathbf{Z}_{6}$ \\ \hline
\textbf{\ \ \ 7} & \textbf{\ +1} & \textbf{\ \ \ --1} & $\mathbf{\ \ \ Z}_{7}%
\mathbf{Z}_{8}\mathbf{Z}_{9}$ \\ \hline
\textbf{\ \ \ 8} & \textbf{\ +1} & \textbf{\ \ \ --1} & $\mathbf{\ \ \ Z}_{7}%
\mathbf{Z}_{8}\mathbf{Z}_{9}$ \\ \hline
\textbf{\ \ \ 9} & \textbf{\ +1} & \textbf{\ \ \ --1} & $\mathbf{\ \ \ Z}_{7}%
\mathbf{Z}_{8}\mathbf{Z}_{9}$ \\ \hline
\end{tabular}

\bigskip \textbf{Table 2: The generators's action on a phase flip affected
qubit.}

\bigskip \bigskip

\noindent \noindent {\Large 3.3.7\ Simulation results\ }\bigskip

The Feynman program decribed in {\large [5][6][7][8] }is a set of
procedures\ supporting the definition and manipulation of\ the states\ of an
n-qubits system and\ the unitary gates acting on them. \noindent We\ have
simulated\ the Shor\ code\ on\ Maple 11 using Feynman program as library.
The appendix D contains this code with some included explanations. As the
algorithm structure aims to reduce the running time, we avoided the several
matrix product, because it takes too much time. It contains four parts:
coding , error, correction and decoding, giving respectively, the states
Psi1, Psi2, Psi2b, and Psi3.\bigskip \emph{\ \ \ \ \ \ \ \ \ \ \ \ }

\noindent {\Large \ \ \ }The{\Large \ }table 3 gives the simulation results
when no correction is made before or after decoding. The first column
contains the X$_{\text{i}}$ , Z$_{\text{i}}$ and Y$_{\text{i }}$input errors
on the qubit "i" and the second one the obtained output error. We note that,
the first qubit state is always recovered without correction, at the end of
the quantum circuit. Therefore, we can always suppress the ancillas without
affecting the useful information.\bigskip \noindent\

Also, we also note that, apart X$_{\text{8}}$ and X$_{\text{9}}$, all the
other X$_{\text{i}}$ input errors are corrected at the output. We remark
that, all the Z$_{\text{i}}$ input errors are transformed in X$_{\text{i}}$
output errors by the decoding circuit. Moreover, the number of affected
qubits at the output is one or two for the X$_{\text{i}}$ and Z$_{\text{i}}$
input errors and two, three or four for the Y$_{\text{i}}$ input errors.
Finally, we note that every Y$_{\text{i}}$ input error gives a distinct
output error, which then allows to identify clearly this input error.
Inversely, the same output errors can be produced by different X$_{\text{i}}$
or Z$_{\text{i}}$ input errors.\bigskip

\ \ \ \ \ $\ \
\begin{tabular}{|c|c|}
\hline
Input & Output \\ \hline
$\ \ X_{\text{1}}$ & $X_{\text{4}}X_{\text{5}}$ \\ \hline
$\ X_{\text{2}}$ & $X_{\text{4}}$ \\ \hline
$\ \ X_{\text{3}}$ & $X_{\text{5}}$ \\ \hline
$\ \ X_{\text{4}}$ & $X_{\text{6}}X_{\text{7}}$ \\ \hline
$\ \ X_{\text{5}}$ & $X_{\text{6}}$ \\ \hline
$\ \ X_{\text{6}}$ & $\ X_{\text{7}}$ \\ \hline
$\ \ X_{\text{7}}$ & $X_{\text{8}}X_{\text{9}}$ \\ \hline
$\ \ X_{\text{8}}$ & $X_{\text{8}}$ \\ \hline
$\ \ X_{\text{9}}$ & $X_{\text{9}}$ \\ \hline
\end{tabular}%
\ \ \mathbf{\ }%
\begin{tabular}{|c|c|}
\hline
Input & Output \\ \hline
$Z_{\text{1}}$ & $X_{\text{2}}X_{\text{3}}$ \\ \hline
$Z_{\text{2}}$ & $X_{\text{2}}X_{\text{3}}$ \\ \hline
$Z_{\text{3}}$ & $X_{\text{2}}X_{\text{3}}$ \\ \hline
$Z_{\text{4}}$ & $X_{\text{2}}$ \\ \hline
$Z_{\text{5}}$ & $X_{\text{2}}$ \\ \hline
$Z_{\text{6}}$ & $X_{\text{2}}$ \\ \hline
$Z_{\text{7}}$ & $X_{\text{3}}\ $ \\ \hline
$Z_{\text{8}}$ & $X_{\text{3}}$ \\ \hline
$Z_{\text{9}}$ & $X_{\text{3}}$ \\ \hline
\end{tabular}%
\ \ \mathbf{\ }%
\begin{tabular}{|l|l|}
\hline
Input & \ \ \ Output \\ \hline
$\ \ \ Y_{\text{1}}$ & $-iX_{\text{2}}X_{\text{3}}X_{\text{4}}X_{\text{5}}$
\\ \hline
$\ \ \ Y_{\text{2}}$ & $\ \ -iX_{\text{2}}X_{\text{3}}X_{\text{4}}$ \\ \hline
$\ \ \ Y_{\text{3}}$ & $\ \ -iX_{\text{2}}X_{\text{3}}X_{\text{5}}$ \\ \hline
$\ \ \ Y_{\text{4}}$ & $\ \ -iX_{\text{2}}X_{\text{6}}X_{\text{7}}$ \\ \hline
$\ \ \ Y_{\text{5}}$ & $\ \ \ \ -iX_{\text{2}}X_{\text{6}}$ \\ \hline
$\ \ \ Y_{\text{6}}$ & $\ \ \ \ -iX_{\text{2}}X_{\text{7}}$ \\ \hline
$\ \ \ Y_{\text{7}}$ & $\ \ -iX_{\text{3}}X_{\text{8}}X_{\text{9}}$ \\ \hline
$\ \ \ Y_{\text{8}}$ & $\ \ \ \ -iX_{\text{3}}X_{\text{8}}$ \\ \hline
$\ \ \ Y_{\text{9}}$ & $\ \ \ \ -iX_{\text{3}}X_{\text{9}}$ \\ \hline
\end{tabular}%
\bigskip $\ \ \ \ \

\ \textbf{Table 3: The output errors for a bit and phase flip on one
qubit.\bigskip\ }\bigskip \bigskip

\noindent {\Large 4 Correction of double errors }

\bigskip

Consider now double errors occurring before decoding and corrected by single
error having same syndrome. Tables $5$ give for each input error${\small \ }E%
{\small ,\ }$the syndromes $S$ and error ${\normalsize E}_{i}$ affecting the
protected qubit i=1 after correction and decoding.\bigskip\ \ {\small \ \ \
\ \ \ \ \ }

\ \ \ \ \ \ \ \ \ \ \ \
\begin{tabular}{|l|l|l|l|l|l|}
\hline
${\small \ \ \ \ \ \ E\ }$ & ${\small \ \ \ \ \ \ S}$ & ${\small \ \ E}_{i}$
& ${\small \ \ \ \ \ \ \ \ \ \ E\ }$ & ${\small \ \ \ \ \ \ S}$ & ${\small E}%
_{i}$ \\ \hline
${\small X}_{^{_{1}}}{\small ,X}_{^{2}}{\small X}_{^{3}}$ & ${\small 10000000%
}$ & ${\small I}_{i}{\small ,Z}_{i}$ & ${\small Y}_{1}{\small ,X_{1}Z}_{2,3}$
& ${\small 10000010}$ & ${\small I}_{i}$ \\ \hline
${\small X}_{2}{\small ,X}_{1}{\small X}_{3}$ & ${\small 11000000}$ & $%
{\small I}_{i}{\small ,Z}_{i}$ & ${\small Y}_{2}{\small ,X}_{2}{\small Z}%
_{1,3}$ & ${\small 11000010}$ & ${\small I}_{i}$ \\ \hline
${\small X}_{3}{\small ,X}_{1}{\small X}_{2}$ & ${\small 01000000}$ & $%
{\small I}_{i}{\small ,Z}_{i}{\small \ }$ & ${\small Y}_{3}{\small ,X_{3}Z}%
_{1,2}$ & ${\small 01000010\ }$ & ${\small I}_{i}$ \\ \hline
${\small X}_{4}{\small ,X}_{^{5}}{\small X}_{^{6}}$ & ${\small 00100000}$ & $%
{\small I}_{i}{\small ,Z}_{i}$ & ${\small Y}_{4}{\small ,X}_{4}{\small Z}%
_{5,6}$ & ${\small 00100011}$ & ${\small I}_{i}$ \\ \hline
${\small X}_{5}{\small ,X}_{4}{\small X}_{6}$ & ${\small 00110000}$ & $%
{\small I}_{i}{\small ,Z}_{i}$ & ${\small Y}_{5}{\small ,X}_{5}{\small Z}%
_{4,6}$ & ${\small 00110011}$ & ${\small I}_{i}$ \\ \hline
${\small X}_{6},{\small X}_{4}{\small X}_{5}$ & ${\small 00010000}$ & $%
{\small I}_{i}{\small ,Z}_{i}$ & ${\small Y}_{6}{\small ,X_{6}Z}_{4,5}$ & $%
{\small 00010011}$ & ${\small I}_{i}$ \\ \hline
${\small X}_{7}{\small ,X}_{8}{\small X}_{^{_{_{9}}}}$ & ${\small 00001000}$
& ${\small I}_{i}{\small ,Z}_{i}$ & ${\small Y}_{7}{\small ,X}_{7}{\small Z}%
_{8,9}$ & ${\small 00001001}$ & ${\small I}_{i}$ \\ \hline
${\small X}_{8}{\small ,X_{^{_{7}}}X}_{^{9}}$ & ${\small 00001100}$ & $%
{\small I}_{i}{\small ,Z}_{i}$ & ${\small Y}_{8}{\small ,X}_{8}{\small Z}%
_{7,9}$ & ${\small 00001101}$ & ${\small I}_{i}$ \\ \hline
${\small X}_{9}{\small ,X}_{^{_{7}}}{\small X}_{_{8}}$ & ${\small 00000100}$
& ${\small I}_{i}{\small ,Z}_{i}$ & ${\small Y}_{9}{\small ,X}_{9}{\small Z}%
_{7,8}$ & ${\small 00000101}$ & ${\small I}_{i}$ \\ \hline
\end{tabular}%
\ \ \ \ \ \ \ \ \ \ \ \ \ \ \ \ \ \ \ \ \ \ \ \ \ \ \ \ \ \ \ \ \ \ \ \ \ \
\ \ \ \ \ \ \ \ \ \ \ \ \ \ \ \ \ \ \ \ \ \ \ \ \ \ \ \ \ \ \ \ \ \ \ \ \ \
\ \ \ \ \ \ \ \ \ \ \ \ \ \ \ \ \ \ \ \ \ \ \ \ \ \ \ \ \ \ \ \ \ \ \ \ \ \
\ \ \ \ \ \ \ \ \ \ \ \ \ \ \ \ \ \ \ \ \ \ \ \ \ \ \ \ \ \ \ \ \ \ \ \ \ \
{\Large \ \ }

\noindent

\noindent

\bigskip

\noindent\textbf{Table }$\mathbf{4a:}$\textbf{\ }Syndromes $S$ and the error
$E_{i}$ $\ $affecting the protected qubits $"i=1"$ after correction by an
operators ${\small \ X}_{k}$ or ${\small Y}_{k\text{ }}$ $(k=1..9).\ \ \ $

\noindent\ \ \ \ \ \ \ \ \ \ \ \
\begin{tabular}{|l|l|l|}
\hline
${\small \ \ \ \ \ \ \ \ \ \ \ \ \ \ \ \ \ \ \ \ \ \ \ \ \ \ \ Errors}$ & $%
{\small \ \ \ \ \ \ S}$ & $\ {\small E}_{i}$ \\ \hline
$({\small Z}_{1}{\small ,Z}_{2}{\small ,Z}_{3}),({\small Z}_{4}{\small %
Z_{_{^{7,8,9}}},Z}_{^{5}}{\small Z}_{^{_{_{_{{\small 7,8,9}}}}}}{\small ,Z}%
_{_{^{6}}}{\small Z}_{_{^{_{7,8,9}}}})$ & ${\small 00000010}$ & $({\small I}%
_{i}),({\small X}_{i})$ \\ \hline
$({\small Z}_{4}{\small ,Z}_{5}{\small ,Z}_{6}),({\small Z}_{^{_{1}}}{\small %
Z}_{^{_{7,8,9}}}{\small ,Z}_{^{_{2}}}{\small Z}_{^{_{7,8,9}}}{\small ,Z}%
_{^{3}}{\small Z}_{^{7,8,9}})$ & ${\small 00000011}$ & $({\small I}_{i}),(%
{\small X}_{i})$ \\ \hline
$({\small Z}_{7}{\small ,Z}_{8}{\small ,Z}_{9}),{\small (Z}_{^{_{1}}}{\small %
Z}_{{\small 4,5,6}}{\small ,Z}_{^{_{2}}}{\small Z}_{{\small 4,5,6}}{\small ,Z%
}_{^{3}}{\small Z}_{{\small 4,5,6}}{\small )\ }$ & ${\small 00000001}$ & $(%
{\small I}_{i}),({\small X}_{i})$ \\ \hline
$({\small Z}_{^{1}}{\small Z_{^{_{2,3}}},Z_{^{2}}Z_{^{3}},Z}_{^{4}}{\small Z}%
_{^{5,6}}{\small ,Z}_{^{5}}{\small Z}_{^{6}}{\small ,Z}_{^{7}}{\small Z}%
_{^{8,9}}{\small ,Z}_{^{8}}{\small Z}_{^{9}})$ & ${\small 00000000}$ & $\ \
{\small (I}_{i}{\small )}$ \\ \hline
\end{tabular}%
\bigskip

\noindent \textbf{Table }$\mathbf{4b}$\textbf{: }{\small Syndromes }$S$%
{\small \ and error }$E_{i}${\small \ on the protected qubits }$"i=1"$%
{\small \ after correction by an operator }$\ {\small Z}_{k}.${\small %
\bigskip }\ \ \

\ \ \ \ \ \ \ \ \ \
\begin{tabular}{|l|l|l|l|l|l|}
\hline
$\ \ {\small E\ }$ & ${\small \ \ \ \ \ \ S}$ & $\ \ {\small E\ }$ & $%
{\small \ \ \ \ \ \ S}$ & $\ {\small E\ }$ & ${\small \ \ \ \ \ \ S}$ \\
\hline
${\small \ X}_{^{_{1}}}{\small X}_{4}$ & ${\small 10100000}$ & ${\small X}%
_{^{_{2}}}{\small X}_{7}$ & ${\small 11001000}$ & ${\small X}_{^{_{4}}}%
{\small X}_{7}$ & ${\small 00101000}$ \\ \hline
${\small \ X}_{^{_{1}}}{\small X}_{5}$ & ${\small 10110000}$ & ${\small X}%
_{^{_{2}}}{\small X}_{8}$ & ${\small 11001100}$ & ${\small X}_{^{_{4}}}%
{\small X}_{8}$ & ${\small 00101100}$ \\ \hline
${\small \ X}_{^{_{1}}}{\small X}_{6}$ & ${\small 10010000}$ & ${\small X}%
_{^{_{2}}}{\small X}_{9}$ & ${\small 11000100}$ & ${\small X}_{^{4}}{\small X%
}_{^{9}}$ & ${\small 00100100}$ \\ \hline
${\small \ X}_{^{_{1}}}{\small X}_{7}$ & ${\small 10001000}$ & ${\small X}%
_{^{3}}{\small X}_{^{4}}$ & ${\small 01100000}$ & ${\small X}_{^{5}}{\small X%
}_{7}$ & ${\small 00111000}$ \\ \hline
${\small \ X}_{^{_{1}}}{\small X}_{8}$ & ${\small 10001100}$ & ${\small X}%
_{^{3}}{\small X}_{^{5}}$ & ${\small 01110000}$ & ${\small X}_{^{5}}{\small X%
}_{8}$ & ${\small 00111100}$ \\ \hline
${\small \ X}_{^{_{1}}}{\small X}_{9}$ & ${\small 10000100}$ & ${\small X}%
_{^{3}}{\small X}_{^{6}}$ & ${\small 01010000}$ & ${\small X}_{^{5}}{\small X%
}_{9}$ & ${\small 00110100}$ \\ \hline
${\small \ X}_{^{2}}{\small X}_{^{4}}$ & ${\small 11100000}$ & ${\small X}%
_{^{3}}{\small X}_{7}$ & ${\small 01001000}$ & ${\small X}_{6}{\small X}_{7}$
& ${\small 00011000}$ \\ \hline
${\small \ X}_{^{2}}{\small X}_{^{5}}$ & ${\small 11110000}$ & ${\small X}%
_{^{3}}{\small X}_{8}$ & ${\small 01001100}$ & ${\small X}_{^{6}}{\small X}%
_{^{8}}$ & ${\small 00011100}$ \\ \hline
${\small \ X}_{^{_{2}}}{\small X}_{6}$ & ${\small 11010000}$ & ${\small X}%
_{3}{\small X}_{9}$ & ${\small 01000100}$ & ${\small X}_{^{_{6}}}{\small X}%
_{_{9}}$ & ${\small 00010100}$ \\ \hline
\end{tabular}%
\bigskip

\noindent \textbf{Table }$\mathbf{4c:}$\textbf{\ }{\small Syndromes }$S$%
{\small \ of double channels errors }$X_{k}X_{l}${\small \ not affecting the
protected qubits after correction}$.${\small \bigskip }\ \ \ \ \ \ \ \ \ \ \
\ \ \ \ \ \ \ \ \ \

\bigskip

\ \ \ \ \ \ \ \ \ \ \ \ \ \ \ \ \ \ \
\begin{tabular}{|l|l|l|l|}
\hline
${\small \ \ \ \ E\ }$ & ${\small \ \ \ \ \ \ S}$ & ${\small \ \ \ \ \ E\ }$
& ${\small \ \ \ \ \ \ S}$ \\ \hline
${\small X}_{1}{\small Z}_{{\small 4,5,6}}$ & ${\small 10000011}$ & ${\small %
X}_{{\small 4}}{\small Z}_{1,2,3}{\small \ }$ & ${\small 00100010}$ \\ \hline
${\small X}_{1}{\small Z}_{{\small 7,8,9}}$ & ${\small 10000001}$ & ${\small %
X}_{{\small 5}}{\small Z}_{1,2,3}{\small \ }$ & ${\small 00110010}$ \\ \hline
${\small X}_{2}{\small Z}_{{\small 7,8,9}}{\small \ }$ & ${\small 11000001}$
& ${\small X}_{{\small 6}}{\small Z}_{1,2,3}{\small \ }$ & ${\small 00010010}
$ \\ \hline
${\small X}_{2}{\small Z}_{{\small 4,5,6}}$ & ${\small 11000011}$ & ${\small %
X}_{{\small 7}}{\small Z}_{1,2,3}$ & ${\small 00001010}$ \\ \hline
${\small X}_{3}{\small Z}_{{\small 4,5,6}}{\small \ }$ & ${\small 01000011}$
& ${\small X}_{{\small 8}}{\small Z}_{1,2,3}$ & ${\small 00001110}$ \\ \hline
${\small X}_{3}{\small Z}_{{\small 7,8,9}}{\small \ }$ & ${\small 01000001}$
& ${\small X}_{{\small 9}}{\small Z}_{1,2,3}{\small \ }$ & ${\small 00000110}
$ \\ \hline
${\small X}_{{\small 4}}{\small Z}_{{\small 7,8,9}}{\small \ }$ & ${\small %
00100001}$ & ${\small X}_{{\small 7}}{\small Z}_{4,5,6}$ & ${\small 00001011}
$ \\ \hline
${\small X}_{{\small 5}}{\small Z}_{{\small 7,8,9}}$ & ${\small 00110001}$ &
${\small X}_{{\small 8}}{\small Z}_{4,5,6}$ & ${\small 00001111}$ \\ \hline
${\small X}_{{\small 6}}{\small Z}_{{\small 7,8,9}}$ & ${\small 00010001}$ &
${\small X}_{{\small 9}}{\small Z}_{4,5,6}$ & ${\small 00000111}$ \\ \hline
\end{tabular}%
\bigskip

\noindent \textbf{Table }$\mathbf{4d}$\textbf{: }Syndromes of double
channels errors $X_{k}Z_{l}$ not affecting the protected qubits after
correction$.$\bigskip \bigskip

\noindent {\Large 5 The seven qubits code}

\bigskip

This code described in {\normalsize [9][10],} uses seven qubits to protect
one of them in a superposed state from any error X, Y or Z. The tables $6$
give syndromes and error ${\normalsize E}_{i}$ affecting the protected
qubits "$i=1$" (after correction) for different single and double channels
errors. The generators of this code are\ : $\ {\small g}_{1}{\small =X}_{%
{\small 4}}{\small X}_{{\small 5}}{\small X}_{{\small 6}}{\small X}_{{\small %
7}}{\small ,\ \ g}_{2}{\small =X}_{{\small 2}}{\small X}_{{\small 3}}{\small %
X}_{{\small 6}}{\small X}_{{\small 7}}{\small ,\ }$

${\small \ g}_{3}{\small =X}_{{\small 1}}{\small X}_{{\small 3}}{\small X}_{%
{\small 5}}{\small X}_{{\small 7}}{\small ,\ \ \ g}_{4}{\small =Z}_{{\small 4%
}}{\small Z}_{{\small 5}}{\small Z}_{{\small 6}}{\small Z}_{{\small 7}}%
{\small ,}$\noindent \noindent ${\small \ }$\noindent ${\small g}_{5}{\small %
=Z}_{{\small 2}}{\small Z}_{{\small 3}}{\small Z}_{{\small 6}}{\small Z}_{%
{\small 7}}{\small ,\ \ g}_{6}{\small =Z}_{{\small 1}}{\small Z}_{{\small 3}}%
{\small Z}_{{\small 5}}{\small Z}_{{\small 7}}$.\ \ \ \ \ \ \ \ \ \ \ \ \ \
\ \ \ \ \ \ \ \ \ \ \ \ \ \ \ \ \ \ \ \ \ \ \ \ \ \ \ \ \ \ \ \ \ \ \ \ \ \
\ \ \ \ \ \ \ \ \ \ \ \ \ \ \ \ \ \ \ \ \ \ \ \ \ \ \ \ \ \ \ \ \ \ \ \ \ \
\ \ \ \ \ \ \ \ \ \ \ \ \ \ \ \ \ \ \ \ \ \ \ \ \ \ \ \ \ \ \ \ \ \ \ \ \ \
\ \ \ \ \ \ \ \ \ \ \ \ \ \ \ \ \ \ \ \ \ \ \ \ \ \ \ \ \ \ \ \ \ \ \ \ \ \
\ \ \ \ \ \ \ \ \ \ \ \ \ \ \ \ \ \ \ \ \ \ \ \ \ \ \ \ \ \ \ \ \ \ \ \ \ \
\ \ \ \ \ \ \ \ \ \ \ \ \ \ \ \ \ \ \ \ \ \ \ \ \ \ \ \ \ \ \ \ \ \ \ \ \ \
\ \ \ \ \ \ \ \ \ \ \ \ \ \ \ \ \ \ \ \ \ \ \ \ \ \ \ \ \ \ \ \ \ \ \ \ \ \
\ \ \ \ \ \ \ \ \ \ \ \ \ \ \ \ \ \ \ \ \ \ \ \ \ \ \ \ \ \ \ \ \ \ \ \ \ \
\ \ \ \ \ \ \ \ \ \ \ \ \ \ \ \ \ \ \ \ \ \ \ \ \ \ \ \ \ \ \ \ \ \ \ \ \ \
\ \ \ \ \ \ \ \ \ \ \ \ \ \ \ \ \ \ \ \ \ \ \ \ \ \ \ \ \ \ \ \ \ \ \ \ \ \
\ \ \ \ \ \ \ \ \ \ \ \ \ \ \ \ \ \ \ \ \ \ \ \ \ \ \ \ \ \ \ \ \ \ \ \ \ \
\ \ \ \ \ \ \ \ \ \ \ \ \ \ \ \ \ \ \ \ \ \ \ \ \ \ \ \ \ \ \ \ \ \ \ \ \ \
\ \ \ \ \ \ \ \ \ \ \ \ \ \ \ \ \ \ \ \ \ \ \ \ \ \ \ \ \ \ \ \ \ \ \ \ \ \
\ \ \ \ \ \ \ \ \ \ \ \ \ \ \ \ \ \ \ \ \ \ \ \ \ \ \ \ \ \ \ \ \ \ \ \ \ \
\ \ \ \ \ \ \ \ \ \ \ \ \ \ \ \ \ \ \ \ \ \ \ \ \ \ \ \ \ \ \ \ \ \ \ \ \ \
\ \ \ \ \ \ \ \ \ \ \ \ \ \ \ \ \ \ \ \ \ \ \ \ \ \ \ \ \ \ \ \ \ \ \ \ \ \
\ \ \ \ \ \ \ \ \ \ \ \ \ \ \ \ \ \ \ \ \ \ \ \ \ \ \ \ \ \ \ \ \ \ \ \ \ \
\ \ \ \ \ \ \ \ \ \ \ \ \ \ \ \ \ \ \ \ \ \ \ \ \ \ \ \ \ \ \ \ \ \ \ \ \ \
\ \ \ \ \ \ \ \ \ \ \ \ \ \ \ \ \ \ \ \ \ \ \ \ \ \ \ \ \ \ \ \ \ \ \ \ \ \
\ \ \ \ \ \ \ \ \ \ \ \ \ \ \ \ \ \ \ \ \ \ \ \ \ \ \ \ \ \ \ \ \ \ \ \ \ \
\ \ \ \ \ \ \ \ \ \ \ \ \ \ \ \ \ \ \ \ \ \ \ \ \ \ \ \ \ \ \ \ \ \ \ \ \ \
\ \ \ \ \ \ \ \ \ \ \ \ \ \ \ \ \ \ \ \ \ \ \ \ \ \ \ \ \ \ \ \ \ \ \ \ \ \
\ \ \ \ \ \ \ \ \ \ \ \ \ \ \ \ \ \ \ \ \ \ \ \ \ \ \ \ \ \ \ \ \ \ \ \ \ \
\ \ \ \ \ \ \ \ \ \ \ \ \ \ \ \ \ \ \ \ \ \ \ \ \ \ \ \ \ \ \ \ \ \ \ \ \ \
\ \ \ \ \ \ \ \ \ \ \ \ \ \ \ \ \ \ \ \ \ \ \ \ \ \ \ \ \ \ \ \ \ \ \ \ \ \
\ \ \ \ \ \ \ \ \ \ \ \ \ \ \ \ \ \ \ \ \ \ \ \ \ \ \ \ \ \ \ \ \ \ \ \ \ \
\ \ \ \ \ \ \ \ \ \ \ \ \ \ \ \ \ \ \ \ \ \ \ \ \ \ \ \ \ \ \ \ \ \ \ \ \ \
\ \ \ \ \ \ \ \ \ \ \ \ \ \ \ \ \ \ \ \ \ \ \ \ \ \ \ \ \ \ \ \ \ \ \ \ \ \
\ \ \ \ \ \ \ \ \ \ \ \ \ \ \ \ \ \ \ \ \ \ \ \ \ \ \ \ \ \ \ \ \ \ \ \ \ \
\ \ \ \ \ \ \ \ \ \ \ \ \ \ \ \ \ \ \ \ \ \ \ \ \ \ \ \ \ \ \ \ \ \ \ \ \ \
\ \ \ \ \ \ \ \ \ \ \ \ \ \ \ \ \ \ \ \ \ \ \ \ \ \ \ \ \ \ \ \ \ \ \ \ \ \
\ \ \ \ \ \ \ \ \ \ \ \ \ \ \ \ \ \ \ \ \ \ \ \ \ \ \ \ \ \ \ \ \ \ \ \ \ \
\ \ \ \ \ \ \ \ \ \ \ \ \ \ \ \ \ \ \ \ \ \ \ \ \ \ \ \ \ \ \ \ \ \ \ \ \ \
\ \ \ \ \ \ \ \ \ \ \ \ \ \ \ \ \ \ \ \ \ \ \ \ \ \ \ \ \ \ \ \ \ \ \ \ \ \
\ \ \ \ \ \ \ \ \ \ \ \ \ \ \ \ \ \ \ \ \ \ \ \ \ \ \ \ \ \ \ \ \ \ \ \ \ \
\ \ \ \ \ \ \ \ \ \ \ \ \ \ \ \ \ \ \ \ \ \ \ \ \ \ \ \ \ \ \ \ \ \ \ \ \ \
\ \ \ \ \ \ \ \ \ \ \ \ \ \ \ \ \ \ \ \ \ \ \ \ \ \ \ \ \ \ \ \ \ \ \ \ \ \
\ \ \ \ \ \ \ \ \ \ \ \ \ \ \ \ \ \ \ \ \ \ \ \ \ \ \ \ \ \ \ \ \ \ \ \ \ \
\ \ \ \ \ \ \ \ \ \ \ \ \ \ \ \ \ \ \ \ \ \ \ \ \ \ \ \ \ \ \ \ \ \ \ \ \ \
\ \ \ \ \ \ \ \ \ \ \ \ \ \ \ \ \ \ \ \ \ \ \ \ \ \ \ \ \ \ \ \ \ \ \ \ \ \
\ \ \ \ \ \ \ \ \ \ \ \ \ \ \ \ \ \ \ \ \ \ \ \ \ \ \ \ \ \ \ \ \ \ \ \ \ \
\ \ \ \ \ \ \ \ \ \ \ \ \ \ \ \ \ \ \ \ \ \ \ \ \ \ \ \ \ \ \ \ \ \ \ \ \ \
\ \ \ \ \ \ \ \ \ \ \ \ \ \ \ \ \ \ \ \ \ \ \ \ \ \ \ \ \ \ \ \

\noindent

\ \
\begin{tabular}{|l|l|l|l|l|l|}
\hline
$\ \ \ \ \ \ \ \ \ \ \ \ \ \ {\small E}$ & ${\small \ \ \ \ \ S}$ & ${\small %
\ \ \ \ E}_{i}$ & ${\small E}$ & ${\small \ \ \ \ \ S}$ & ${\small E}_{i}$
\\ \hline
${\small X}_{1},({\small X}_{2}{\small X}_{3},{\small X_{4}X}_{{\small 5}},%
{\small X_{6}X}_{7})$ & ${\small 000001}$ & ${\small I}_{i}{\small ,(X}_{i})%
{\small \ }$ & ${\small Y}_{1}$ & ${\small 001001}$ & ${\small I}_{i}$ \\
\hline
${\small X}_{2},({\small X}_{1}{\small X}_{3},{\small X_{4}X}_{6},{\small X}%
_{{\small 5}}{\small X}_{7})$ & ${\small 000010}$ & ${\small I}_{i},{\small %
(X}_{i}){\small \ }$ & ${\small Y}_{2}{\small \ }$ & ${\small 010010}$ & $%
{\small I}_{i}$ \\ \hline
${\small X}_{3},({\small X}_{1}{\small X}_{2},{\small X}_{{\small 5}}{\small %
X}_{6},{\small X_{4}X}_{7})$ & ${\small 000011}$ & ${\small I}_{i}{\small %
,(X_{i})\ \ }$ & ${\small Y}_{3}$ & ${\small 011011\ }$ & ${\small I}_{i}$
\\ \hline
${\small X}_{4},({\small X}_{1}{\small X}_{5},{\small X}_{2}{\small X}_{6},%
{\small X}_{3}{\small X}_{7})$ & ${\small 000100}$ & ${\small I}_{i}{\small %
,(X_{i})\ }$ & ${\small Y}_{4}{\small \ }$ & ${\small 100100}$ & ${\small I}%
_{i}{\small \ }$ \\ \hline
${\small X}_{5},({\small X}_{1}{\small X}_{4},{\small X}_{2}{\small X}_{7},%
{\small X}_{3}{\small X}_{^{6}})$ & ${\small 000101}$ & ${\small I}_{i}%
{\small ,(X_{i})\ }$ & ${\small Y}_{5}$ & ${\small 101101}$ & ${\small I}%
_{i} $ \\ \hline
${\small X}_{6},({\small X}_{1}{\small X}_{7},{\small X}_{2}{\small X_{4}},%
{\small X}_{3}{\small X}_{{\small 5}})$ & ${\small 000110}$ & ${\small I}%
_{i},{\small (X}_{i})$ & ${\small Y}_{6}$ & ${\small 110110}$ & ${\small I}%
_{i}$ \\ \hline
${\small X}_{7},({\small X}_{1}{\small X}_{6},{\small X}_{2}{\small X}_{%
{\small 5}},{\small X}_{3}{\small X_{4}})$ & ${\small 000111}$ & ${\small I}%
_{i}{\small ,(X_{i})\ }$ & ${\small Y}_{7}$ & ${\small 111111}$ & ${\small I}%
_{i}$ \\ \hline
\end{tabular}
\bigskip

\noindent \textbf{Table }$\mathbf{5a}$\textbf{: }{\small Syndromes and error
}$E_{i}${\small \ on the protected qubits "i" after correction by an
operators}$\ {\small X}_{k}$ or ${\small Y}_{k}.$\bigskip

\ \ \ \ \ \ \ \ \ \ \ \ \ \ \ \ \ \ \ \
\begin{tabular}{|l|l|l|}
\hline
$\ \ \ \ \ \ \ \ Error$ & ${\small \ \ \ \ S}$ & $\ {\small E}_{i}$ \\ \hline
${\small Z_{^{_{1}}},(Z_{^{_{_{_{6}}}}}Z_{_{^{7}}}},{\small Z_{^{2}}Z}%
_{^{3}},{\small Z_{^{4}}Z}_{^{5}}{\small )}$ & ${\small 001000}$ & ${\small I%
}_{i},{\small (Z}_{i}){\small \ }$ \\ \hline
${\small Z}_{{\small 2}},({\small Z_{^{1}}Z}_{^{3}},{\small Z_{^{4}}Z}%
_{^{6}},{\small Z_{^{5}}Z}_{^{7}})$ & ${\small 010000}$ & ${\small I}_{i},%
{\small (Z}_{i}){\small \ }$ \\ \hline
${\small Z}_{3},({\small Z_{^{4}}Z}_{^{7}},{\small Z_{^{1}}Z}_{^{2}},{\small %
Z_{^{5}}Z}_{^{6}})$ & ${\small 011000}$ & ${\small I}_{i},{\small (Z}_{i})%
{\small \ }$ \\ \hline
${\small Z}_{4},({\small Z_{^{3}}Z}_{^{7}},{\small Z_{^{1}}Z}_{^{5}},{\small %
Z_{^{2}}Z}_{^{6}})$ & ${\small 100000}$ & ${\small I}_{i},{\small (Z}_{i})%
{\small \ }$ \\ \hline
${\small Z}_{5},({\small Z_{^{2}}Z_{^{7}},Z_{^{_{1}}}Z}_{4},{\small %
Z_{^{_{3}}}Z}_{^{6}})$ & ${\small 101000}$ & ${\small I}_{i},{\small (Z}_{i})%
{\small \ \ }$ \\ \hline
${\small Z}_{6},({\small Z_{^{1}}Z}_{^{7}},{\small Z_{^{2}}Z}_{^{4}},{\small %
Z_{^{3}}Z}_{^{5}})$ & ${\small 110000}$ & ${\small I}_{i},{\small (Z}_{i})%
{\small \ }$ \\ \hline
${\small Z}_{7},({\small Z_{^{1}}Z}_{^{6}},{\small Z_{^{2}}Z}_{^{5}},{\small %
Z_{^{3}}Z}_{^{4}})$ & ${\small 111000}$ & ${\small I}_{i},{\small (Z}_{i})%
{\small \ \ }$ \\ \hline
\end{tabular}%
\bigskip

\noindent \textbf{Table }$\mathbf{5b}$ \textbf{: }{\small Syndromes and the
error affecting the protected qubits after correction by an operator}$\
{\small Z}_{k}.${\small \bigskip }$\ $\ \ \ \ \ \ \ \ \

\ \ \ \ \ \ \ \ \ \ \
\begin{tabular}{|l|l|l|l|l|l|}
\hline
${\small Errors}$ & ${\small \ \ \ \ S}$ & ${\small Errors}$ & ${\small \ \
\ \ \ S}$ & ${\small Errors}$ & ${\small \ \ \ \ S}$ \\ \hline
${\small X}_{1}{\small Z}_{4}$ & ${\small 100001}$ & ${\small Z}_{1}{\small X%
}_{4}{\small \ }$ & ${\small 001100}$ & ${\small Z}_{1}{\small X}_{3}$ & $%
{\small 001011}$ \\ \hline
${\small X}_{1}{\small Z}_{5}$ & ${\small 101001}$ & ${\small Z}_{2}{\small X%
}_{4}$ & ${\small 010100}$ & ${\small X}_{^{_{1}}}{\small Z}_{2}$ & ${\small %
010001}$ \\ \hline
${\small X}_{1}{\small Z}_{6}$ & ${\small 110001}$ & ${\small Z}_{3}{\small X%
}_{4}$ & ${\small 011100}$ & ${\small X}_{^{_{1}}}{\small Z}_{3}{\small \ }$
& ${\small 011001}$ \\ \hline
${\small X}_{1}{\small Z}_{7}$ & ${\small 111001}$ & ${\small Z}_{1}{\small X%
}_{5}$ & ${\small 001101}$ & ${\small X}_{2}{\small Z}_{1}{\small \ \ }$ & $%
{\small 001010}$ \\ \hline
${\small X}_{2}{\small Z}_{7}$ & ${\small 111010}$ & ${\small Z}_{2}{\small X%
}_{5}$ & ${\small 010101}$ & ${\small X}_{2}{\small Z}_{3}{\small \ }$ & $%
{\small 011010}$ \\ \hline
${\small X}_{3}{\small Z}_{4}{\small \ }$ & ${\small 100011}$ & ${\small Z}%
_{1}{\small X}_{6}$ & ${\small 001110}$ & ${\small X}_{2}{\small Z}_{4}$ & $%
{\small 100010\ }$ \\ \hline
${\small X}_{3}{\small Z}_{6}$ & ${\small 110011}$ & ${\small Z}_{2}{\small X%
}_{6}{\small \ }$ & ${\small 010110}$ & ${\small X}_{2}{\small Z}_{5}{\small %
\ }$ & ${\small 101010}$ \\ \hline
${\small X}_{3}{\small Z}_{7}{\small \ }$ & ${\small 111011}$ & ${\small Z}%
_{3}{\small X}_{6}$ & ${\small 011110}$ & ${\small X}_{2}{\small Z}_{6}%
{\small \ \ }$ & ${\small 110010}$ \\ \hline
${\small X}_{4}{\small Z}_{7}{\small \ \ }$ & ${\small 111100}$ & ${\small Z}%
_{1}{\small X}_{7}$ & ${\small 001111}$ & ${\small Z}_{2}{\small X}_{3}$ & $%
{\small 010011}$ \\ \hline
${\small X}_{5}{\small Z}_{7}$ & ${\small 111101}$ & ${\small Z}_{2}{\small X%
}_{7}$ & ${\small 010111}$ & ${\small X}_{4}{\small Z}_{5}$ & ${\small 101100%
}$ \\ \hline
${\small X}_{6}{\small Z}_{5}$ & ${\small 101110}$ & ${\small Z}_{3}{\small X%
}_{7}{\small \ }$ & ${\small 011111}$ & ${\small X}_{4}{\small Z}_{6}$ & $%
{\small 110100}$ \\ \hline
${\small X}_{6}{\small Z}_{7}$ & ${\small 111110}$ & ${\small Z}_{4}{\small X%
}_{5}$ & ${\small 100101}$ & ${\small Z}_{4}{\small X}_{6}$ & ${\small 100110%
}$ \\ \hline
${\small X}_{5}{\small Z}_{6}$ & ${\small 110101}$ & ${\small Z}_{3}{\small X%
}_{5}$ & ${\small 011101}$ & ${\small X}_{3}{\small Z}_{5}$ & ${\small 101011%
}$ \\ \hline
\end{tabular}%
\bigskip

\noindent \textbf{Table }$\mathbf{5c}$\textbf{: }{\small Syndromes of double
channels errors }${\small Z}_{k}{\small X}_{l}${\small \ not affecting the
protected qubits after correction.}

\bigskip \noindent

\bigskip

\noindent {\Large 6 The five qubits code}\noindent

\bigskip

This code is described in {\normalsize [11]} and uses five qubits to protect
one of them in a superposed state from any error X, Y or Z. The four
stabilizers of this code are:$\bigskip $

$g_{1}=XXZIZ\ \ \ \ g_{2}=ZXXZI$ \ \ \ \ \ $g_{3}=IZXXZ$ $\ \ \ \
g_{4}=ZIZXX\ \ $\bigskip

The tables $7$\ give the syndromes S of single and double errors occurring
in the transmitting channel. The double errors are corrected as the single
error having same syndrome. The third column gives the error $E_{i}$
affecting after decoding the first to be protected qubit "$i=1$" $\left\vert
\Psi _{i}\right\rangle =\alpha _{i}\left\vert 0\right\rangle +\beta
_{i}\left\vert 1\right\rangle .$

\bigskip\ \ \ \ \ \ \ \ \ \ \ \ \ \ \ \ \
\begin{tabular}{|l|l|l|}
\hline
${\small \ \ \ \ \ \ \ \ \ \ \ \ \ \ \ \ \ \ \ Error}$ & ${\small \ \ \ S}$
& ${\small \ \ \ \ \ \ \ E}_{i}$ \\ \hline
${\small X}_{i}{\small ,(Z}_{3}{\small Z}_{4}{\small ),(X}_{4}{\small Z}%
_{_{5}}{\small ,Z}_{2}{\small X}_{_{3}}{\small )}$ & ${\small 0101}$ & $%
{\small I}_{i}{\small ,(X}_{i}{\small ),(Z}_{i}{\small )}$ \\ \hline
${\small X}_{2}{\small ,(Z}_{4}{\small Z}_{_{5}}{\small ),(Z}_{i}{\small X}%
_{_{5}}{\small ,Z}_{3}{\small X}_{_{4}}{\small )}$ & ${\small 0010}$ & $%
{\small I}_{i}{\small ,(X}_{i}{\small ),(Z}_{i}{\small )}$ \\ \hline
${\small X}_{3}{\small ,(Z}_{i}{\small Z}_{_{5}}{\small ),(X}_{i}{\small Z}%
_{2}{\small ,Z}_{4}{\small X}_{5}{\small )}$ & ${\small 1001}$ & ${\small I}%
_{i}{\small ,(X}_{i}{\small ),(Z}_{i}{\small )}$ \\ \hline
${\small X}_{4}{\small ,(Z}_{i}{\small Z}_{2}{\small ),(X}_{i}{\small Z}_{5}%
{\small ,X}_{2}{\small Z}_{3}{\small )}$ & ${\small 0100}$ & ${\small I}_{i}%
{\small ,(X}_{i}{\small ),(Z}_{i}{\small )}$ \\ \hline
${\small X}_{5}{\small ,(Z}_{2}{\small Z}_{_{3}}{\small ),(X}_{3}{\small Z}%
_{4}{\small ,Z}_{i}{\small X}_{2}{\small )}$ & ${\small 1010}$ & ${\small I}%
_{i}{\small ,(X}_{i}{\small ),(Z}_{i}{\small )}$ \\ \hline
${\small Z}_{i}{\small ,(X}_{2}{\small X}_{_{5}}{\small ),(X}_{3}{\small Z}%
_{_{5}}{\small ,Z}_{2}{\small X}_{4}{\small )}$ & ${\small 1000}$ & ${\small %
I}_{i}{\small ,(X}_{i}{\small ),(Z}_{i}{\small )}$ \\ \hline
${\small Z}_{2}{\small ,(X}_{i}{\small X}_{3}{\small ),(Z}_{i}{\small X}_{4}%
{\small ,Z}_{3}{\small X}_{_{5}}{\small )}$ & ${\small 1100}$ & ${\small I}%
_{i}{\small ,(X}_{i}{\small ),(Z}_{i}{\small )}$ \\ \hline
${\small Z}_{3}{\small ,(X}_{2}{\small X}_{_{4}}{\small ),(X}_{i}{\small Z}%
_{_{4}}{\small ,Z}_{2}{\small X}_{5}{\small )}$ & ${\small 0110}$ & ${\small %
I}_{i}{\small ,(X}_{i}{\small ),(Z}_{i}{\small )}$ \\ \hline
${\small Z}_{4}{\small ,(X}_{3}{\small X}_{5}{\small ),(X}_{i}{\small Z}_{3}%
{\small ,X}_{2}{\small Z}_{_{5}}{\small )}$ & ${\small 0011}$ & ${\small I}%
_{i}{\small ,(X}_{i}{\small ),(Z}_{i}{\small )}$ \\ \hline
${\small Z}_{5}{\small ,(X}_{i}{\small X}_{_{4}}{\small ),(X}_{2}{\small Z}%
_{_{4}}{\small ,Z}_{i}{\small X}_{3}{\small )}$ & ${\small 0001}$ & ${\small %
I}_{i}{\small ,(X}_{i}{\small ),(Z}_{i}{\small )}$ \\ \hline
\end{tabular}

\noindent

\bigskip

\noindent\textbf{\noindent Table} $\mathbf{6a}$ : Double errors corrected as
$X$ or $Z$ single errors having same syndrome.\noindent\ The ancillas are
designed by $"a_{j}"$ and the to be protected qubit by $"i"$ with $i=1$ and\
$\ j=1,2,3,4.$

\ \ \ \ \ \ \ \ \ \ \ \ \ \ \ \ \ \ \ \ \ \ \ \ \ \
\begin{tabular}{|l|l|l|}
\hline
{\small \ \ \ \ \ \ }${\small \ \ \ \ Error}$ & ${\small \ \ \ S}$ & $%
{\small \ \ \ E}_{i}$ \\ \hline
${\small Y}_{i}{\small ,(X}_{3}{\small X}_{_{4}}{\small ,Z}_{2}{\small Z}%
_{_{5}}{\small )}$ & ${\small 1101}$ & ${\small I}_{i}{\small ,(Y}_{i}%
{\small )}$ \\ \hline
${\small Y}_{2}{\small ,(X}_{4}{\small X}_{_{5}}{\small ,Z}_{i}{\small Z}_{3}%
{\small )}$ & ${\small 1110}$ & ${\small I}_{i}{\small ,(Y}_{i}{\small )}$
\\ \hline
${\small Y}_{3}{\small ,(X}_{i}{\small X}_{5}{\small ,Z}_{2}{\small Z}_{4}%
{\small )}$ & ${\small 1111}$ & ${\small I}_{i}{\small ,(Y}_{i}{\small )}$
\\ \hline
${\small Y}_{4}{\small ,(X}_{i}{\small X}_{2}{\small ,Z}_{3}{\small Z}_{_{5}}%
{\small )}$ & ${\small 0111}$ & ${\small I}_{i}{\small ,(Y}_{i}{\small )}$
\\ \hline
${\small Y}_{5}{\small ,(X}_{2}{\small X}_{_{3}}{\small ,Z}_{i}{\small Z}%
_{_{4}}{\small )}$ & ${\small 1011}$ & ${\small I}_{i}{\small ,(Y}_{i}%
{\small )}$ \\ \hline
\end{tabular}%
\bigskip

\noindent \textbf{Table }$\mathbf{6b}$ :\ Double errors $X_{k}X_{l}$ and $%
Z_{k}Z_{l}$ corrected as $Y$ errors having same syndrome.\noindent\

\bigskip

\noindent

\bigskip

\noindent{\Large 7 Fidelity}\bigskip

The fidelity $F({\large \sigma },{\large \rho })$ is one of the mathematical
quantities which permits to know how close are two quantum states
represented by the density matrix $\sigma $ and $\rho ,$ by measuring a
distance between them{\large \ }$[1]${\large \ : }$\bigskip $

$\ \ \ \ \ \ \ \ \ \ \ \ \ \ \ \ \ \ \ \ \ \ \ \ \ \ \ \ \ \ \ \ \ \ \ \ F(%
{\large \sigma },{\large \rho })=\left\vert Tr(\sqrt{\sqrt{\sigma }\rho
\sqrt{\sigma }})\right\vert ^{2}$ \ \ \ \ \ \ \ \ \ \ \ \ \ \ \ \ \ \ \ \ \
\ \ \ \ \ \ \ $(19)$

\bigskip

In the case of a pure state $\sigma =\left\vert \Psi \right\rangle
\left\langle \Psi \right\vert $\ and an arbitrary state $\rho $, the
fidelity is the overlap between them $[1]$\ :{\large \ }\bigskip\ $\ \ \ \ \
\ \ \ \ \ \ \ \ \ \ \ \ \ \ \ \ \ \ \ \ \ \ \ \ \ \ \ \ \ \ \ \ \ \ \ \ $

$\ \ \ \ \ \ \ \ \ \ \ \ \ \ \ \ \ \ \ \ \ \ \ \ \ \ \ \ \ \ \ \ \ \ \ \ \ \
\ F(\left\vert \Psi \right\rangle ,{\large \rho })=\left\langle \Psi
\right\vert {\large \rho }\left\vert \Psi \right\rangle $ \ \ \ \ \ \ \ \ \
\ \ \ \ \ \ \ \ \ \ \ \ \ \ \ \ \ \ \ \ \ \ \ \ $(20)$

\bigskip

In this work we measure the overlap between the useful qubit input state $%
\sigma =\left\vert \Psi \right\rangle \left\langle \Psi \right\vert $\ and
the output state $\rho =\left\vert \Psi ^{E}\right\rangle \left\langle \Psi
^{E}\right\vert $ affected by channel error E and obtained after decoding.
Then, the fidelity is function of the angles $(\theta ,\phi )$ in the Block
sphere and the average fidelity is : $\bigskip \noindent $

$\ \ \ \ \ F_{a}=(1/4\pi )\diint F(\theta ,\phi )sin(\theta )d\theta d\phi ,$%
\ \ with $0\leq \theta \leq \pi $ and $0\leq \phi \leq 2\pi $\ \ \ $\ \ \ \
(21)$

\bigskip

Consider any double channel error $E_{k}E_{l}$ occurring on the logical
qubit $"ia"$ (useful qubit $"i"$ protected by ancillas $"a"$) during the
transmission through a depolarizing channel and corrected as the single
error with similar syndrome. Suppose $P$ the probability that any single
channel error $E_{i}$ occurs on qubit $"i".$\ Then, the global density
matrix of the system received is :\bigskip

\noindent\ $\ \ \ \ \ \ \ \ \ \ \ \ \ \ \ \ \ \ \ \ \ \ \ \rho
_{_{ia}}^{E}=(1-P)^{2}\rho _{_{_{ia}}}+P(1-P)(\rho _{_{ia}}^{E_{k}}+\rho
_{_{ia}}^{E_{l}})+P^{2}\rho _{_{ia}}^{E_{k}E_{l}}$\ $\ \ \ \ \ \ \ \ \ \ \ \
\ \ \ \ \ \ \ \ \ (22)$

\bigskip

With $\rho _{_{_{ia}}},$ $\rho _{_{ia}}^{E_{k}},$ $\rho _{_{ia}}^{E_{l}}$
and $\rho _{_{ia}}^{E_{k}E_{l}}$ the density matrix respectively unaffected
and affected by $E_{k},$ $E_{l}$ and $E_{k}E_{l}$ errors.\bigskip\

After decoding and suppressing the ancillas we obtain the matrix density of
the useful qubit:\bigskip

$\ \ \ \ \ \ \ \ \ \ \ \ \ \ \ \ \ \ \ \ \ \rho _{_{i}}^{E}=(1-P)^{2}\rho
_{_{_{i}}}+P(1-P)(\rho _{_{i}}+\rho _{_{i}})+P^{2}\rho _{_{i}}^{E_{i}}$ $\ \
\ \ \ \ \ \ \ \ \ \ \ \ \ \ \ \ \ \ \ \ \ \ (23)$\bigskip

We note that the single errors $E_{k}$ and $E_{l}$ are recovered by the
three codes and only the double error $E_{k}E_{l}$ will affect the protected
qubits by error $E_{i}=$ $I_{i},$ $X_{i},$ $Y_{i}$ or $Z_{i}.$ Then we
obtain :\bigskip

$\ \ \ \ \ \ \ \ \ \ \ \ \ \ \ \ \ \ \ \ \ \ \ \ \ \ \ \ \ \ \ \ \ \ \rho
_{_{i}}^{E}\ =(1-P^{2})\rho _{_{i}}+P^{2}\rho _{_{i}}^{E_{i}}$ \ \ \ \ \ \ \
\ \ \ \ \ \ \ \ \ \ \ \ \ \ \ \ \ \ \ \ \ \ \ \ \ $(24)$\bigskip

We multiply by the initial state $\left\vert {\small \Psi }_{i}\right\rangle
={\small \alpha }_{i}\left\vert {\small 0}\right\rangle {\small +\beta }%
_{i}\left\vert {\small 1}\right\rangle ${\small = }${\small cos(\theta /2)}%
\left\vert {\small 0}\right\rangle +{\small e}^{i\phi }{\small sin}({\small %
\theta /2})\left\vert {\small 1}\right\rangle $ and obtain:$\bigskip $

$\ \ \ \ \ \ \ \ \ \ \ \ \ \ \ \ \ \ \ \ \left\langle \Psi _{i}\right\vert
\rho _{_{i}}^{E}\left\vert \Psi _{i}\right\rangle =(1-P^{2})\left\langle
\Psi _{i}\right\vert \rho _{_{i}}\left\vert \Psi _{i}\right\rangle
+P^{2}\left\langle \Psi _{i}\right\vert \rho _{_{i}}^{E_{i}}\left\vert \Psi
_{i}\right\rangle $ \ \ \ \ \ \ \ $\ \ \ \ \ (25)$\bigskip

\ \ \ \ \ \ \ \ \ \ \ \ \ \ \ \ \ \ \ \ \ With $\left\langle \Psi
_{i}\right\vert \rho _{_{i}}\left\vert \Psi _{i}\right\rangle =1$ and $%
F^{E_{i}}(\theta ,\phi )$=$\left\langle \Psi _{i}\right\vert \rho
_{i}^{E_{i}}\left\vert \Psi _{i}\right\rangle \ \ \ \ \ \ \ \ \ \ \ \ \ \ \
\ \ \ (26)$\bigskip \bigskip

The fidelity (overlap) ${\small F(P,\theta ,\phi )=}\left\langle {\small %
\Psi }_{i}\right\vert {\small \rho }_{_{i}}^{{\small E}}\left\vert {\small %
\Psi }_{i}\right\rangle $ is then :\bigskip $\bigskip $

$\ \ \ \ \ \ \ \ \ \ \ \ \ \ \ \ \ \ \ \ \ \ \ \ \ \ \ F(P,\theta ,\phi
)=(1-P^{2})+P^{2}F^{E_{i}}(\theta ,\phi )$ \ \ \ $\ \ \ \ \ \ \ \ \ \ \ \ \
\ \ \ \ \ \ \ \ \ \ \ \ \ \ \ \ \ \ (27)$\bigskip

Finally we obtain the average fidelity :

$\bigskip $

$\ \ \ \ \ \ \ \ \ \ \ \ \ \ {\small F}_{a}{\small (P)=(}\frac{{\small 1}}{%
{\small 4\pi }}{\small )}\diint {\small F(P,\theta ,\phi )sin(\theta
)d\theta d\phi }$ $({\small 0\leq \theta }${\small \ }${\small \leq \pi }$ $%
{\small ,0\leq \phi }$ ${\small \leq 2\pi })$ \ \ \ \ \ \ $(28)$\bigskip
\bigskip \bigskip

$\ \ \ \ \ \ \ \ \ \ \ \ \ \ \ \ \ \ \ \ \ \ \
F_{a}(P)=(1-P^{2})+P^{2}F_{a}^{E_{i}}$ $=1+P^{2}(\ F_{a}^{E_{i}}-1)\ \ \ \ \
\ \ \ \ \ \ \ \ \ \ \ \ \ \ \ \ \ \ \ \ \ \ \ \ (29)$\bigskip \bigskip
\bigskip

Table 8 gives the fidelity and average fidelity for errors $E_{i}$ occurring
on the useful qubit "i" with probability P=1.\bigskip\ \ \ \ \ \ \ \ \ \ \ \
\ \ \ \ \ \ \ \ \ \ \ \ \ \ \ \ \ \ \ \ \ \ \noindent\ \ \ \ \ \ \ \ \ \ \ \
\ \ \ \ \

\ \ \ \ \ \ \ \ \ \ \ \ \ \ \ \ \ \ \ \ \ \ \ \ \ \ \ \ \ \ \
\begin{tabular}{|l|l|l|}
\hline
${\small E}_{i}{\small \ }$ & ${\small \ \ \ F}^{E_{i}}{\small (\theta ,\phi
)\ }$ & $F_{a}$ \\ \hline
${\small X}_{i}$ & $\left\vert {\small sin(\theta )cos\phi }\right\vert ^{%
{\small 2}}$ & ${\small 1/3}$ \\ \hline
${\small Y}_{i}{\small \ }$ & $\left\vert {\small sin(\theta )sin\phi }%
\right\vert ^{{\small 2}}$ & ${\small 1/3}$ \\ \hline
${\small Z}_{i}$ & ${\small \ \ \ \ cos}^{2}{\small (\theta )}$ & ${\small %
1/3}$ \\ \hline
\end{tabular}

\noindent \textbf{Table }$\mathbf{7}$ \textbf{:} Fidelity\textbf{\ }${\small %
F(\theta ,\phi )=}\left\vert {\small Tr}(\sqrt{\sqrt{{\small \sigma }_{i}}%
{\small \rho }_{i}^{{\small E}}\sqrt{{\small \sigma }_{i}}})\right\vert ^{2}%
{\small =}\left\langle {\small \Psi }_{i}\right\vert {\small (}\left\vert
{\small \Psi }_{{\small i}}^{{\small E}}\right\rangle \left\langle {\small %
\Psi }_{{\small i}}^{{\small E}}\right\vert {\small )}\left\vert {\small %
\Psi }_{i}\right\rangle $ versus errors on the useful qubits $"i".$\bigskip

\bigskip

\noindent{\Large 8 Comparison among the three codes}

\bigskip

Let us now compare the three codes by computing the average fidelity when
double errors occur in a depolarizing channel. \ For this purpose, we define
for any code C$_{n}$ an average value $f_{n}$ for the parameter $F_{a}$
depicted in table 7 :\bigskip

$\ \ \ \ \ \ \ \ \ \ \ \ \ \ \ \ \ \ \ \ \ \ \ \ \ \ \ \ \ \ \ \ \ \ \ \ \ \
f_{n}=\frac{[x_{n}\times 1+(N_{n}-x_{n})\times \frac{1}{3}]}{N_{n}}$ \ \ \ \
\ \ \ \ \ \ $\ \ \ \ \ \ \ \ \ \ \ \ \ \ \ \ \ \ \ \ \ \ \ \ \ \ \ \ \ \ \ \
\ (30)$\bigskip \bigskip \bigskip

With $x_{n}$ the number of double errors letting the useful qubit unaffected
and $N_{n}$ their total number.$\bigskip \ $From tables 5, 6 and 7 we obtain
for the three codes :$\ \ \ \ \ \ \ \ \ \ \ \ \ \ \ \ \ \ \ \ \ \ \ $

$\ \ \ \ \ \ \ \ \ \ \ \ \ \ \ \ \ \ \ \ \ \ \ f_{5}=\frac{1}{3}=\frac{27}{81%
}$ \ \ \ ; \ \ $f_{7}=\frac{53}{81}$ \ \ \ ; \ \ \ \ $f_{9}=\frac{120}{144}=%
\frac{5}{6}\bigskip $ \ \ \ \ \ \ $\ \ \ \ \ \ \ \ \ \ \ \ (31)$\bigskip
\bigskip \bigskip

If the useful qubit "$i$" is sent without protection through a depolarizing
channel it will be affected by error $E_{i}=X_{i},$ $Y_{i}$ or $Z_{i}$
error, then the affected matrix density received is :\bigskip

$\ \ \ \ \ \ \ \ \ \ \ \ \ \ \ \ \ \ \ \ \ \ \ \ \ \ \ \ \ \ \ \ \ \ \ \
\rho _{_{i}}^{E}=(1-P)\rho _{_{_{i}}}+P\rho _{_{i}}^{E_{i}}$\ $\ \ \ \ \ \ \
\ \ \ \ \ \ \ \ \ \ \ \ \ \ \ \ \ \ \ \ \ \ \ \ \ \ (32)$\bigskip

Then the average fidelity:\bigskip

$\ \ \ \ \ \ \ \ \ \ \ \ \ \ \ \ \ \ \ \ \ \ \ \ \ \ \ \ \ \ \ \ \
F_{a}(P)=(1-P)+PF_{a}^{E_{i}}$ $=1-\frac{2}{3}P\ \ \ \ \ \ \ \ \ \ \ \ \ \ \
\ \ \ \ \ \ \ \ \ (33)$\bigskip

With $F_{a}^{E_{i}}=\frac{1}{3}$ for $E_{i}=X_{i},$ $Y_{i}$ or $Z_{i}$%
\bigskip

The table 8 depicts the expression of average fidelity for each
code.\bigskip\

\ \ \ \ \ \ \ \
\begin{tabular}{|l|l|l|l|l|l|}
\hline
Code & $\ \ C_{0}$ & $\ \ \ \ \ \ \ \ C_{n}$ & $\ \ \ \ C_{5}$ & $\ \ \ \
C_{7}$ & $\ \ \ \ C_{9}$ \\ \hline
$F_{a}(P)$ & $1-\frac{2}{3}P$ & $1+P^{2}(f_{n}-1)$ & $1-\frac{2}{3}P^{2}$ & $%
1-\frac{28}{81}P^{2}$ & $1-\frac{1}{6}P^{2}\ $ \\ \hline
\end{tabular}%
\ \ \ \ \ \ \ \ \ \ \ \ \ \ \ \ \ \ \ \ \

\bigskip

\noindent

\noindent\textbf{Table 8 }: The average fidelity calculated for the n-qubits
codes, when double errors occur in a depolarizing channel. The symbols $%
C_{0} $ and $C_{n}$ represent respectively no protection and an $({\small n},%
{\small k=1})$ code. \bigskip

The figure 5 shows that the nine qubits code gives the best fidelity average
fidelity $F_{a}(P)$ for all values of P , followed by the seven then the
five qubits codes.%

\noindent \textbf{Figure} $5$ : The average fidelity with and without
correction. $\ $\bigskip \noindent\

\noindent{\Large 5. Conclusion \ \ }

{\large \ \ \ \ \ }\ \ \ \ \ \ \ \ \ \ \ \ \ \ \ \ \ \ \ \ \ \ \ \ \ \ \ \ \
\ \ \ \ \ \ \ \ \ \ \ \ \ \ \ \ \ \ \ \ \ \ \ \ \ \ \ \ \ \ \ \ \ \ \ \ \ \
\ \ \ \ \ \ \ \ \ \ \ \ \ \ \ \ \ \ \ \ \ \ \ \ \ \ \ \ \ \ \ \ \ \ \ \ \ \
\ \ \ \ \ \ \ \ \ \ \ \ \ \ \ \ \ \ \ \ \ \ \ \ \ \ \ \ \ \ \ \ \ \ \ \ \ \
\ \ \ \ \ \ \ \ \ \ \ \ \ \ \ \ \ \ \ \ \ \ \ \ \ \ \ \ \ \ \ \ \ \ \ \ \ \
\ \ \ \ \ \ \ \ \ \

This work is the first attempt to use Feynman program to simulate quantum
correction. Our results show that this program allows to manipulate easily
the qubits states and the operators acting on them. The running time is low
when we act on a single qubit but increases with the number of qubits on
which the gates act. We have verified that the output are correct for an X,
Z or Y input error on the first qubit.\ Also, the simulation allows us to
conclude that the nine qubits code gives the best average fidelity, followed
by the seven then the five qubits code, regardless the depolarizing channel
error probability P$.$ We have considered triple errors and more very
unlikely and then with negligible effect on the obtained results.\bigskip\

{\large Acknowledgement}\ \ \ \

\ \ \ \ \ \ \ \ \ \ \ \ \ \ \ \ \ \ \ \ \ \ \ \ \ \ \ \ \ \ \ \ \ \ \ \ \ \
\ \ \ \ \ \ \ \ \ \ \ \ \ \ \ \ \ \ \ \ \ \ \ \ \ \ \ \ \ \ \ \ \ \ \ \ \ \
\ \ \ \ \ \ \ \ \ \ \ \ \ \ \ \ \ \ \ \ \ \ \ \ \ \ \ \ \ \ \ \ \ \ \ \ \ \
\ \ \ \ \ \ \ \ \ \ \ \ \ \ \ \ \ \ \ \ \ \ \ \ \ \ \ \ \ \ \ \ \ \ \ \ \ \
\ \ \ \ \ \ \ \ \ \ \ \ \ \ \ \ \ \ \ \ \ \ \ \ \ \ \ \ \ \ \ \ \ \ \ \ \ \
\ \ \ \ \ \ \ \ \ \ \ \ \ \ \ \ \ \ \ \ \ \ \ \ \ \ \ \ \ \ \ \ \ \ \ \ \ \
\ \ \ \ \ \ \ \ \ \ \ \ \ \ \ \ \ \ \ \ \ \ \ \ \ \ \ \ \ \ \ \ \ \ \ \ \ \
\ \ \ \ \ \ \ \ \ \ \ \ \ \ \ \ \ \ \ \ \ \ \ \ \ \ \ \ \ \ \ \ \ \ \ \ \ \
\noindent\ \ \ \ \ \ \ \ \ \ \ \

We would like to thank very much Mrs. S.Fritzsch and T.Radtke for providing
us with the version 4\ (2008) of Feynman Program.

\noindent

\bigskip \newpage

\noindent {\Large 6.\ References \ \ \ \ }\bigskip

\noindent {\normalsize [1]}\ M.A. Nielsen, I.L Chuang: 'Quantum computation
and quantum information', Cambridge University Press, UK, 2000.\noindent
{\normalsize \ }

\noindent {\normalsize [2] }Daniel Gottesman, "An Introduction to Quantum
Error Correction"\ , Proceeding of Symposium in Applied Mathematics.\
2000.\noindent {\normalsize \ }

\noindent {\normalsize [3]}\ S. J.Lomonaco, Jr. , "A\ Rosetta Stone for
Quantum Mechanics with an Introduction to Quantum Computation", \ Proceeding
of Symposis in Appl Math.\ 2000.

\noindent {\normalsize [4] }P.W.Shor, \ Scheme for reducing decoherence in
quantum computer memory. Phys.Rev. A, \ 52(4):R2493-R2496, Oct 1995. \ \ \ \
\ \ \ \ \ \ \ \ \ \ \ \ \ \ \ \ \ \ \ \ \ \ \ \ \ \ \ \

\noindent {\normalsize [5]}{\large \ }T.Radtke, S.Fritzsche: 'Simulation of
n-qubits quantum systems', I.\ Quantum gates and registers, Computer Physics
Communications, 2005.

\noindent {\normalsize [6]}{\large \ }T.Radtke, S.Fritzsche: 'Simulation of
n-qubits quantum systems', II.\ Quantum states, Computer Physics
Communications, 2006.

\noindent {\normalsize [7]}{\large \ }T.Radtke, S.Fritzsche: 'Simulation of
n-qubits quantum systems', III.\ Quantum operations, Computer Physics
Communications, 2007.

\noindent {\normalsize [8]}{\large \ }T.Radtke, S.Fritzsche: 'Simulation of
n-qubits quantum systems',IV. Parametrizations of quantum states, Computer\
Physics Communications, 2008.\noindent {\normalsize \ }

\noindent {\normalsize [9]} A. M. Steane. Multiple particle interference and
quantum error correction. Proc. R. Soc. Lond. A, 452:2551--2576, 1996.
quant-ph/9601029.

\noindent {\normalsize [10]} Austin G. Fowler, "Constructing arbitrary
Steane code single logical qubit fault-tolerant gates",
AarXiv:quant-ph/0411206v2, 20 Dec 2010.\noindent {\normalsize \ }

\noindent {\normalsize [11]}\textbf{\ }R.Laflamme,1 C.Miquel,1,2 J.Pablo
Paz,1,2 and Wojciech H.Zurek1, "Perfect..Code", Physic Review Letters,
Volume\ 77, Number 1, july 1996.\ \ \ \ \ \ \ \ \ \ \ \ \ \ \ \ \ \ \ \ \ \
\ \ \ \ \ \ \ \ \ \ \ \ \ \ \ \ \ \ \ \ \ \ \ \ \ \ \ \ \ \ \ \ \ \ \ \ \ \
\ \ \ \ \ \ \ \ \ \ \ \ \ \ \ \ \ \ \ \ \ \ \ \ \ \ \ \ \ \ \ \ \ \ \ \ \ \
\ \ \ \ \ \ \ \ \ \ \ \ \ \ \ \ \ \ \ \ \ \ \ \ \ \ \ \ \ \ \ \ \ \ \ \ \ \
\ \ \ \ \ \ \ \ \ \ \ \ \ \ \ \ \ \ \ \ \ \ \ \ \ \ \ \ \ \ \ \ \ \ \ \ \ \
\ \ \ \ \ \ \ \ \ \ \ \ \ \ \ \ \ \ \ \ \ \ \ \ \ \ \ \ \ \ \ \ \ \ \ \ \ \
\ \ \ \ \ \ \ \ \ \ \ \ \ \ \ \ \ \ \ \ \ \ \ \ \ \ \ \ \ \ \ \ \ \ \ \ \ \
\ \ \ \ \ \ \ \ \ \ \ \ \ \ \ \ \ \ \ \ \ \ \ \ \ \ \ \ \ \ \ \ \ \ \ \ \ \
\ \ \ \ \ \ \ \ \ \ \ \ \ \ \ \ \ \ \ \ \ \ \ \ \ \ \ \ \ \ \ \ \ \ \ \ \ \
\ \ \ \ \ \ \ \ \ \ \ \ \ \ \ \ \ \ \ \ \ \ \ \ \ \ \ \ \ \ \ \ \ \ \ \ \ \
\ \ \ \ \ \ \ \ \ \ \ \ \ \ \ \ \ \ \ \ \ \ \ \ \ \ \ \ \ \ \ \ \ \ \ \ \ \
\ \ \ \ \ \ \ \ \ \ \ \ \ \ \ \ \ \ \ \ \ \ \ \ \ \ \ \ \ \ \ \ \ \ \ \ \ \
\ \ \ \ \ \ \ \ \ \ \ \ \ \ \ \ \ \ \ \ \ \ \ \ \ \ \ \ \ \ \ \ \ \ \ \ \ \
\ \ \ \ \ \ \ \ \ \ \ \ \ \ \ \ \ \ \ \ \ \ \ \ \ \ \ \ \ \ \ \ \ \ \ \ \ \
\ \ \ \ \ \ \ \ \ \ \ \ \ \ \ \ \ \ \ \ \ \ \ \ \ \ \ \ \ \ \ \ \ \ \ \ \ \
\ \ \ \ \ \ \ \ \ \ \ \ \ \ \ \ \ \ \ \ \ \ \ \ \ \ \ \ \ \ \ \ \ \ \ \ \ \
\ \ \ \ \ \ \ \ \ \ \ \ \ \ \ \ \ \ \ \ \ \ \ \ \ \ \ \ \ \ \ \ \ \ \ \ \ \
\ \ \ \ \ \ \ \ \ \ \ \ \ \ \ \ \ \ \ \ \ \ \ \ \ \ \ \ \ \ \ \ \ \ \ \ \ \
\ \ \ \ \ \ \ \ \ \ \ \ \ \ \ \ \ \ \ \ \ \ \ \ \ \ \ \ \ \ \ \ \ \ \ \ \ \
\ \ \ \ \ \ \ \ \ \ \ \ \ \ \ \ \ \ \ \ \ \ \ \ \ \ \ \ \ \ \ \ \ \ \ \ \ \
\ \ \ \ \ \ \ \ \ \ \ \ \ \ \ \ \ \ \ \ \ \ \ \ \ \ \ \ \ \ \ \ \ \ \ \ \ \
\ \ \ \ \ \ \ \ \ \ \ \ \ \ \ \ \ \ \ \ \ \ \ \ \ \ \ \ \ \ \ \ \ \ \ \ \ \
\ \ \ \ \ \ \ \ \ \ \ \ \ \ \ \ \ \ \ \ \ \ \ \ \ \ \ \ \ \ \ \ \ \ \ \ \ \
\ \ \ \ \ \ \ \ \ \ \ \ \ \ \ \ \ \ \ \ \ \ \ \ \ \ \ \ \ \ \ \ \ \ \ \ \ \
\ \ \ \ \ \ \ \ \ \ \ \ \ \ \ \ \ \ \ \ \ \ \ \ \ \ \ \ \ \ \ \ \ \ \ \ \ \
\ \ \ \ \ \ \ \ \ \ \ \ \ \ \ \ \ \ \ \ \ \ \ \ \ \ \ \ \ \ \ \ \ \ \ \ \ \
\ \ \ \ \ \ \ \ \ \ \ \ \ \ \ \ \ \ \ \ \ \ \ \ \ \ \ \ \ \ \ \ \ \ \ \ \ \
\ \ \ \ \ \ \ \ \ \ \ \ \ \ \ \ \ \ \ \ \ \ \ \ \ \ \ \ \ \ \ \ \ \ \ \ \ \
\ \ \ \ \ \ \ \ \ \ \ \ \ \ \ \ \ \ \ \ \ \ \ \ \ \ \ \ \ \ \ \ \ \ \ \ \ \
\ \ \ \ \ \ \ \ \ \ \ \ \ \ \ \ \ \ \ \ \ \ \ \ \ \ \ \ \ \ \ \ \ \ \ \ \ \
\ \ \ \ \ \ \ \ \ \ \ \ \ \ \ \ \ \ \ \ \ \ \ \ \ \ \ \ \ \ \ \ \ \ \ \ \ \
\ \ \ \ \ \ \ \ \ \ \ \ \ \ \ \ \ \ \ \ \ \ \ \ \ \ \ \ \ \ \ \ \ \ \ \ \ \
\ \ \ \ \ \ \ \ \ \ \ \ \ \ \ \ \ \ \ \ \ \ \ \ \ \ \ \ \ \ \ \ \ \ \ \ \ \
\ \ \ \ \ \ \ \ \ \ \ \ \ \ \ \ \ \ \ \ \ \ \ \ \ \ \ \ \ \ \ \ \ \ \ \ \ \
\ \ \ \ \ \ \ \ \ \ \ \ \ \ \ \ \ \ \ \ \ \ \ \ \ \ \ \ \ \ \ \ \ \ \ \ \ \
\ \ \ \ \ \ \ \ \ \ \ \ \ \ \ \ \ \ \ \ \ \ \ \ \ \ \ \ \ \ \ \ \ \ \ \ \ \
\ \ \ \ \ \ \ \ \ \ \ \ \ \ \ \ \ \ \ \ \ \ \ \ \ \ \ \ \ \ \ \ \ \ \ \ \ \
\ \ \ \ \ \ \ \ \ \ \ \ \ \ \ \ \ \ \ \ \ \ \ \ \ \ \ \ \ \ \ \ \ \ \ \ \ \
\ \ \ \ \ \ \ \ \ \ \ \ \ \ \ \ \ \ \ \ \ \ \ \ \ \ \ \ \ \ \ \ \ \ \ \ \ \
\ \ \ \ \ \ \ \ \ \ \ \ \ \ \ \ \ \ \ \ \ \ \ \ \ \ \ \ \ \ \ \ \ \ \ \ \ \
\ \ \ \ \ \ \ \ \ \ \ \ \ \ \ \ \ \ \ \ \ \ \ \ \ \ \ \ \ \ \ \ \ \ \ \ \ \
\ \ \ \ \ \ \ \ \ \ \ \ \ \ \ \ \ \ \ \ \ \ \ \ \ \ \ \ \ \ \ \ \ \ \ \ \ \
\ \ \ \ \ \ \ \ \ \ \ \ \ \ \ \ \ \ \ \ \ \ \ \ \ \ \ \ \ \ \ \ \ \ \ \ \ \
\ \ \ \ \ \ \ \ \ \ \ \ \ \ \ \ \ \ \ \ \ \ \ \ \ \ \ \ \ \ \ \ \ \ \

\noindent\ \ \newpage\ {\Large \ \ \ \ \ \ \ \ \ \ \ \ \ \ \ \ \ \ \ \ \ \ \
\ \ \ \ \ \ \ \ \ \ \ \ \ \ \ \ \ \ \ \ \ \ \ \ \ \ \ \ \ \ \ \ \ \ \ \ \ \
\ \ \ \ \ \ \ \ \ \ \ \ \ \ \ \ \ \ \ \ \ \ \ \ \ \ \ \ \ \ \ \ \ \ \ \ \ \
\ \ \ \ \ \ \ \ \ \ \ \ \ \ \ \ \ \ \ \ \ \ \ \ \ \ \ \ \ \ \ \ \ \ \ \ \ \
\ \ \ \ \ \ \ \ \ \ \ \ \ \ \ \ \ \ \ \ \ \ \ \ \ \ \ \ \ \ \ \ \ \ \ \ \ \
\ \ \ \ \ \ \ \ \ \ \ \ \ \ \ \ \ \ \ \ }\noindent \noindent {\Large 6.}
{\Large Appendix : Simulations \ }

\bigskip

\noindent {\Large A. Notations}

{\Large \ \ \ \ \ \ \ \ \ \ \ \ \ \ \ \ \ \ }

\noindent

\noindent \textbf{Shor} := proc (Co, Cx, Cy, Cz, n, k):\textit{\ \ }the
procedure called "Shor" takes as input the amplitude probability Co, Cx, Cy,
Cz for respectively, no error, X, Y and Z error on the n$^{\text{th}}\ $%
qubit. The input k=0 is chosen if we want to detect and correct error before
decoding. This choice is useful if we want to use again the ancillas in
their original state $\left\vert 0\right\rangle $. We choose k=1 if we
suppress the ancillas at the end, after measuring their final states to know
the output error. In fact, this last choose is necessary if an arbitrary
error occurs because the correction procedure does not work in this case.

\noindent \textbf{Psio}\ := Feynman\_evaluate("Kronecker product",
\TEXTsymbol{<}a, b\TEXTsymbol{>}, \TEXTsymbol{<}1, 0\TEXTsymbol{>},
\TEXTsymbol{<}1, 0\TEXTsymbol{>}):\ \ Tensor product of the three first
qubit states \TEXTsymbol{<}a, b\TEXTsymbol{>}, \TEXTsymbol{<}1, 0\TEXTsymbol{%
>} and \TEXTsymbol{<}1, 0\TEXTsymbol{>}.

\noindent \textbf{H}:= Feynman\_quantum\_operator("H"): \ matrix for the
Hadamard gate.

\noindent \textbf{X}:= Feynman\_quantum\_operator("X"): \ matrix for the bit
flip X gate.

\noindent \textbf{CN12} := Feynman\_quantum\_operator(3, "cnot", [1, 2]):
matrix for the cnot gate with qubit 2 as target and 1 as controller.

\noindent \textbf{Ha} := Feynman\_quantum\_operator("HHH"): \ Three matrix H
tensor product.

\noindent \textbf{Id}:=Feynman\_quantum\_operator("IIIIIIIII"): The matrix
identity I application\ on all the qubits states.

\noindent \textbf{A}:= Feynman\_evaluate("Kronecker power", \TEXTsymbol{<}1,
0\TEXTsymbol{>}, 6): \ Six identical states \TEXTsymbol{<}1, 0\TEXTsymbol{>}
tensor product.

\noindent \textbf{Psioc}:= Feynman\_evaluate("Kronecker product", Psiob, A):
\ Application of the operator A on the state Psiob.

\noindent \textbf{P1}:= Feynman\_quantum\_operator("permute", [1, 4, 5, 2,
6, 7, 3, 8, 9]): \ Change the order of the qubits so that, for example, the
qubit 4 will be in the second position. \ \

\noindent \textbf{Xa}:= Feynman\_quantum\_operator(9, "X", [1]):\textit{\ }%
The X gate application on the first qubit in a nine qubits system.

\noindent \textbf{T231}:= Feynman\_quantum\_operator(9, "ccn", [2, 3, 1]): \
The Toffoli gate matrix with qubit 2 and 3 as controllers and 1 as target in
the nine qubits system.

\noindent \textbf{Hb}:= Feynman\_quantum\_operator(9,"HHH", [1, 2, 3]):
Application of H on the first, second and third qubits and the identity
matrix I on the six other ones.

\noindent \textbf{Psi33}:= simplify(Feynman\_print(Psi3)): \ Simplify and
display the state Psi3 \ in the Dirac notation.\bigskip \bigskip

\bigskip

\noindent{\Large B. Bit flip code }

{\Large \ \ \ \ \ \ \ }

{\large \TEXTsymbol{>}}with(Feynman): with(LinearAlgebra):
with(Typesetting): Digits := 20:

Bitflip:=proc (n)

local Psio, CN12,CN13,X,T321,E,Psi1,Psi2, Psi3,Psi33:

{\Large \# \ Initial state}

Psio:=Feynman\_evaluate("Kronecker product",\TEXTsymbol{<}a, b\TEXTsymbol{>}%
, \TEXTsymbol{<}1, 0\TEXTsymbol{>}, \TEXTsymbol{<}1, 0\TEXTsymbol{>}):

{\Large \# Coding}

CN12:=Feynman\_quantum\_operator(3, "cnot", [1, 2]):

CN13:=Feynman\_quantum\_operator(3, "cnot", [1, 3]):

Psi11:=CN12.Psio:

Psi1:=CN13.Psi11:

{\Large \# Error}

X := Feynman\_quantum\_operator("X"):

if n = 0 then E := Feynman\_quantum\_operator("III");

print("No error"); end if:

if n = 1 then E:=Feynman\_quantum\_operator(3,"X",[1]);

print("Error X on first qubit"); end if:

if n=2 then E:= Feynman\_quantum\_operator(3, X",[2]);

print("Error X on second qubit"); end if:

if n=3 then E:=Feynman\_quantum\_operator(3, "X", [3]);

print("Error X on third qubit");end if:

Psi2:= E.Psi1:

{\Large \# Decoding}

T321:= Feynman\_quantum\_operator(3,"ccn",[3,2,1]):

Psi31:=CN13.Psi2:

Psi32:=CN12.Psi31:

Psi3:=T321.Psi32:

Psi33:=Feynman\_print(Psi3);print(Psi3 = Psi33);

end proc;

\bigskip

\bigskip \noindent{\Large C. Phase flip code \ }\bigskip

{\large \TEXTsymbol{>}}with(Feynman): with(LinearAlgebra):
with(Typesetting): Digits:=20:

Phaseflip:=proc(n)

local Psio,Psi1,Psi2,Psi3,Psi33,T,H,Ha,CN12,CN13,Hb,Z,Za:

{\Large \# Initial state}

Psio:=Feynman\_evaluate("Kronecker product",\TEXTsymbol{<}a,b\TEXTsymbol{>},%
\TEXTsymbol{<}1,0\TEXTsymbol{>},\TEXTsymbol{<}1,0\TEXTsymbol{>}):

{\Large \# Encoding }

H:=Feynman\_quantum\_operator("H"):

Ha:=Feynman\_quantum\_operator(3,"H",[1]):

CN12:=Feynman\_quantum\_operator(3,"cnot",[1,2]):

CN13:=Feynman\_quantum\_operator(3,"cnot",[1,3]):

Hb:=Feynman\_quantum\_operator("HHH"):

Psi11:=Ha.Psio:

Psi12:=CN12.Psi11:

Psi13:=CN13.Psi12:

Psi1:=Hb.Psi13:

{\Large \# Error }

Z:=Feynman\_quantum\_operator("Z"):

if n=0 then Za:=Feynman\_quantum\_operator("III"):

Psi2:= Za.Psi1: print("No error"); fi:

if n=1 then \ Za:=Feynman\_quantum\_operator(3,"Z",[1]):

Psi2:=Za.Psi1: print("Phase flip on first qubit"); fi:

if n=2 then Za:=Feynman\_quantum\_operator(3,"Z",[2]):

Psi2:=Za.Psi1: print("Phase flip on second qubit"); fi:

if n=3 then Za:=Feynman\_quantum\_operator(3,"Z",[3]):

Psi2:=Za.Psi1: print("Phase flip on third qubit");fi: \

{\Large \# Decoding }

T:=Feynman\_quantum\_operator(3,"ccn",[3,2,1]):

Psi3:=Ha.T.CN12.CN13.Hb.Psi2:

Psi33:=Feynman\_print(Psi3): print(Psi3=Psi33);

end proc;\bigskip

\noindent{\Large D. Shor code}\bigskip

{\large \TEXTsymbol{>}}with(Feynman): with(LinearAlgebra):
with(Typesetting): Digits := 20:

Shor:=proc(Co,Cx,Cy,Cz,n,k)

local Psio,H,X,Y,Z,X1,Z1,X2,Z2,X3,Z3,X4,Z4,X5,Z5,X6,Z6,X7,Z7,X8,Z8,X9,

X123,X456,X789,Z9,X16,X49,Id,CN,CN1,Psioa,Ha,Psiob,Psioc,Psiod,CN2,

CN3,CN4,CN5,CN6,CN7,E,Ex,Ey,Ez,Psi1,Psi2,Psi1a,Psi2a,Psi2b,Psi2c,Psi2d,

Psi2dd,Hb,Psi3a,Psi3,Psi3b,Psi33,T1,T2,T3,T4,A,P1,P2: \ \

{\Large \# Initial state}

Psio:=Feynman\_evaluate("Kronecker product",\TEXTsymbol{<}a,b\TEXTsymbol{>},%
\TEXTsymbol{<}1,0\TEXTsymbol{>},\TEXTsymbol{<}1,0\TEXTsymbol{>}):

{\Large \# Coding}

H:=Feynman\_quantum\_operator("H"):

CN :=Feynman\_quantum\_operator(3,"cnot",[1,2]):

CN1:=Feynman\_quantum\_operator(3,"cnot",[1,3]):

Psioa1 :=CN.Psio:

Psioa:=CN1.Psioa1:

Ha:=Feynman\_quantum\_operator("HHH"):

Psiob:=Ha.Psioa:

A:=Feynman\_evaluate("Kronecker power",\TEXTsymbol{<}1,0\TEXTsymbol{>},6):

Psioc:=Feynman\_evaluate("Kronecker product",Psiob,A):

P1:=Feynman\_quantum\_operator("permute",[1,4,5,2,6,7,3,8,9]):

Psiod:=P1.Psioc:

CN2:=Feynman\_quantum\_operator(9,"cnot",[1,2]):

CN3:=Feynman\_quantum\_operator(9,"cnot",[1,3]): \

CN4:=Feynman\_quantum\_operator(9,"cnot",[4,5]):

CN5:= Feynman\_quantum\_operator(9,"cnot",[4,6]):

CN6:=Feynman\_quantum\_operator(9,"cnot",[7,8]):

CN7:= Feynman\_quantum\_operator(9,"cnot",[7,9]):

Psi1a1:=CN6.Psiod:

Psi1a2:= CN4.Psi1a1:

Psi1a3:= CN2.Psi1a2:

Psi1a4:=CN7.Psi1a3:

Psi1a5:=CN5.Psi1a4:

Psi1:=CN3.Psi1a5:

{\Large \# Error operators}

X:= Feynman\_quantum\_operator("X"):

Y:= Feynman\_quantum\_operator("Y"):

Z:=Feynman\_quantum\_operator("Z"):

Id:=Feynman\_quantum\_operator("IIIIIIIII"):

if n=0 then print("No error");

Ex:= Id:Ey:= Id:Ez:= Id:end if:

if n=1 then if Cx=1 then print("X error on first qubit");end if:

if Cy=1 then print("Y error on first qubit");end if:

if Cz=1 then print("Z error on first qubit");end if:

if Cx\TEXTsymbol{<}1 and Cy\TEXTsymbol{<}1 and Cz\TEXTsymbol{<}1 then
print("Error on first qubit");end if:

Ex:=Feynman\_quantum\_operator(9,"X",[1]):

Ey:=- I.Feynman\_quantum\_operator(9,"Y",[1])):

Ez:=Feynman\_quantum\_operator(9,"Z",[1]):end if:

if n=2 then if Cx=1 then print("X error on second qubit"):end if:

if Cy=1 then print("Y error on second qubit"):end if:

if Cz=1 then print("Z error on second qubit"):end if:

if Cx\TEXTsymbol{<}1 and Cy\TEXTsymbol{<}1 and Cz\TEXTsymbol{<}1 then
print("Error on second qubit"):end if:

Ex:=Feynman\_quantum\_operator(9,"X",[2]):

Ey:=-I.Feynman\_quantum\_operator(9,"Y",[2])):

Ez:= Feynman\_quantum\_operator(9,"Z",[2]):end if:

if n=3 then if Cx=1 then print("X error on third qubit");end if:

if Cy=1 then print("Y error on third qubit");end if:

if Cz=1 then print("Z error on third qubit");end if:

if Cx\TEXTsymbol{<}1 and Cy\TEXTsymbol{<}1 and Cz\TEXTsymbol{<}1 then
print("Error on third qubit");end if:

Ex:=Feynman\_quantum\_operator(9,"X",[3]):

Ey:=-I.Feynman\_quantum\_operator(9,"Y",[3])):

Ez:=Feynman\_quantum\_operator(9,"Z",[3]):end if:

if n=4 then if Cx=1 then print("X error on forth qubit");end if:

if Cy=1 then print("Y error on forth qubit");end if:

if Cz=1 then print("Z error on forth qubit");end if:

if Cx\TEXTsymbol{<}1 and Cy\TEXTsymbol{<}1 and Cz\TEXTsymbol{<}1 then
print("Error on forth qubit");end if:

Ex:= Feynman\_quantum\_operator(9,"X",[4]):

Ey:=-I.Feynman\_quantum\_operator(9,"Y", [4])):

Ez:= Feynman\_quantum\_operator(9,"Z",[4]):end if:

if n=5 then if Cx=1 then print("X error on fifth qubit");end if:

if Cy=1 then print("Y error on fifth qubit");end if:

if Cz=1 then print("Z error on fifth qubit");end if:

if Cx\TEXTsymbol{<}1 and Cy\TEXTsymbol{<}1 and Cz\TEXTsymbol{<}1 then
print("Error on fifth qubit");end if:

Ex:= Feynman\_quantum\_operator(9,"X",[5]):

Ey:=-I.Feynman\_quantum\_operator(9,"Y",[5])):

Ez:= Feynman\_quantum\_operator(9,"Z",[5]):end if:

if n=6 then if Cx=1 then print("X error on sixth qubit");end if:

if Cy=1 then print("Y error on sixth qubit");end if:

if Cz=1 then print("Z error on sixth qubit");end if:

if Cx\TEXTsymbol{<}1 and Cy\TEXTsymbol{<}1 and Cz\TEXTsymbol{<}1 then
print("Error on sixth qubit");end if:

Ex:= Feynman\_quantum\_operator(9,"X",[6]):

Ey:=-I.Feynman\_quantum\_operator(9,"Y",[6])):

Ez:=Feynman\_quantum\_operator(9,"Z",[6]):end if:

if n=7 then if Cx = 1 then print("X error on seventh qubit");end if:

if Cy=1 then print("Y error on seventh qubit");end if:

if Cz=1 then print("Z error on seventh qubit");end if:

if Cx\TEXTsymbol{<}1 and Cy\TEXTsymbol{<}1 and Cz\TEXTsymbol{<}1 then
print("Error on seventh qubit");end if:

Ex:= Feynman\_quantum\_operator(9,"X",[7]):

Ey:=-I, Feynman\_quantum\_operator(9,"Y",[7])):

Ez:= Feynman\_quantum\_operator(9,"Z",[7]):end if:

if n=8 then if Cx=1 then print("X error on eight qubit");end if:

if Cy=1 then print("Y error on eight qubit");end if:

if Cz=1 then print("Z error on eight qubit");end if:

if Cx\TEXTsymbol{<}1 and Cy\TEXTsymbol{<}1 and Cz\TEXTsymbol{<}1 then
print("Error on eight qubit");end if:

Ex:=Feynman\_quantum\_operator(9, "X", [8]):

Ey:=-I.Feynman\_quantum\_operator(9,"Y",[8])):

Ez:= Feynman\_quantum\_operator(9,"Z",[8]):end if:

if n=9 then if Cx=1 then print("X error on nineth qubit");end if:

if Cy=1 then print("Y error on nineth qubit");end if:

if Cz=1 then print("Z error on nineth qubit");end if:

if Cx\TEXTsymbol{<}1 and Cy\TEXTsymbol{<}1 and Cz\TEXTsymbol{<}1 then
print("Error on nineth qubit");end if:

Ex:=Feynman\_quantum\_operator(9,"X",[9]):

Ey:=-I.Feynman\_quantum\_operator(9,"Y",[9])):

Ez:= Feynman\_quantum\_operator(9,"Z",[9]):end if:

{\Large \# Error}

E:=Co.Id+Cx.Ex+Cy.Ey+Cz.Ez :

Psi2 :=E.Psi1:

if k = 0 then print("Correction before decoding");

{\Large \# Gates for error detection and correction\ }

Z1 := Feynman\_quantum\_operator(9,"Z",[1]):

Z2 := Feynman\_quantum\_operator(9,"Z",[2]):

Z3 := Feynman\_quantum\_operator(9,"Z",[3]):

Z4 := Feynman\_quantum\_operator(9, "Z", [4]):

Z5 := Feynman\_quantum\_operator(9, "Z", [5]):

Z6 := Feynman\_quantum\_operator(9, "Z", [6]):

Z7 := Feynman\_quantum\_operator(9, "Z", [7]):

Z8 := Feynman\_quantum\_operator(9,"Z",[8]):

Z9 := Feynman\_quantum\_operator(9, "Z", [9]):

X1:=Feynman\_quantum\_operator(9, "X", [1]):

X2:=Feynman\_quantum\_operator(9,"X",[2]):

X3:=Feynman\_quantum\_operator(9,"X",[3]):

X4 := Feynman\_quantum\_operator(9, "X", [4]):

X5 := Feynman\_quantum\_operator(9, "X", [5]):

X6:=Feynman\_quantum\_operator(9,"X",[6]):

X7:=Feynman\_quantum\_operator(9,"X",[7]):

X8:=Feynman\_quantum\_operator(9,"X",[8]):

X9:=Feynman\_quantum\_operator(9,"X",[9]):

X123:= Feynman\_quantum\_operator(9,"XXX",[1,2,3]):

X456:= Feynman\_quantum\_operator(9,"XXX",[4,5,6]):

X789:= Feynman\_quantum\_operator(9,"XXX",[7,8,9]):

{\Large \# Bit flip detection and correction\ }

if evalb(Equal(Z1.Z2.Psi2,Psi2)) = true

and evalb(Equal(Z2.Z3.Psi2,Psi2))= true

and evalb(Equal(Z4.Z5.Psi2,Psi2))= true

and evalb(Equal(Z5.Z6.Psi2,Psi2))= true

and evalb(Equal(Z7.Z8.Psi2,Psi2))=true

and evalb(Equal(Z8.Z9.Psi2,Psi2)) = true then

Psi2a:=Psi2:end if:

if evalb(Equal (Z1.Z2.Psi2,Psi2))=false

and evalb(Equal(Z2.Z3.Psi2,Psi2)) = true then

Psi2a:=X1.Psi2:end if:

if evalb(Equal (Z1.Z2.Psi2,Psi2))=false

and evalb(Equal(Z2.Z3.Psi2,Psi2))=false then

Psi2a:=X2.Psi2: end if:

if evalb(Equal (Z1.Z2), Psi2), Psi2)) = true

and evalb(Equal(Z2, Z3), Psi2), Psi2)) = false \

then Psi2a :=X3.Psi2 end if:

if evalb(Equal(Z4.Z5.Psi2, Psi2)) = false

and evalb(Equal(Z5.Z6.Psi2, Psi2)) = true then

Psi2a :=X4.Psi2:end if:

if evalb(Equal(Z4.Z5.Psi2, Psi2))=false

and evalb(Equal (Z5.Z6.Psi2, Psi2) = false then

Psi2a :=X5.Psi2: end if:

if evalb(Equal(Z4.Z5.Psi2), Psi2)) = true

and evalb(Equal (Z5.Z6.Psi2, Psi2) = false then

Psi2a :=X6.Psi2: end if:

if evalb(Equal(Z7.Z8.Psi2,Psi2))=false

and evalb(Equal(Z8.Z9,Psi2,Psi2))=true then

Psi2a:=X7.Psi2:end if:

if evalb(Equal(Z7.Z8.Psi2,Psi2))=false

and evalb(Equal(Z8.Z9.Psi2,Psi2)) = false then

Psi2a:=X8.Psi2: end if: if evalb(Equal(Z7.Z8.Psi2, Psi2))=true

and evalb(Equal(Z8.Z9.Psi2,Psi2))=false then

Psi2a:=X9.Psi2:end if:

{\Large \# Phase flip detection and correction\ }

if evalb(Equal(X162.X161.Psi2a,Psi2a))=true

and evalb(Equal(X492.X491.Psi2a,Psi2a))=true then

Psi2b:=Psi2a:end if:

if evalb(Equal(X162.X161.Psi2a,Psi2a))=false

and evalb(Equal(X492.X491.Psi2a, Psi2a)) = true then

Psi2b1:=Z3.Psi2a:

Psi2b2:=Z2.Psi2b1:

Psi2b:=Z1.Psi2b2: end if:

if evalb(Equal(X162.X161.Psi2a, Psi2a))=false

and evalb(Equal(X492.X491.Psi2a,Psi2a))=false then

Psi2b1:=Z6.Psi2a:

Psi2b2:=Z5.Psi2b1:

Psi2b:=Z4.Psi2b2: end if:

if evalb(Equal(X162.X161.Psi2a, Psi2a))=true

and evalb(Equal(X492.X491.Psi2a,Psi2a))=false then

Psi2b1:=Z9.Psi2a:

Psi2b2:=Z8.Psi2b1:

Psi2b:=Z7.Psi2b2: end if:

end if:

if k=1 then print("Decoding without correction");

Psi2b:=Psi2:end if:

{\Large \# Decoding}

T1:=Feynman\_quantum\_operator(9,"ccn",[2,3,1]):

T2:=Feynman\_quantum\_operator(9,"ccn",[5,6,4]):

T3:=Feynman\_quantum\_operator(9,"ccn",[8,9,7]):

Psi2c1:=CN6.Psi2b:

Psi2c2:=CN4.Psi2c1:

Psi2c:=CN2.Psi2c2:

Psi2d1:=CN7.Psi2c :

Psi2d2:=CN5.Psi2d1 :

Psi2d3:=CN3.Psi2d2 :

Psi2d4:=T3.Psi2d3 :

Psi2d5:=T2.Psi2d4 :

Psi2d=T1.Psi2d5 :

P2:= Feynman\_quantum\_operator("permute", [1,4,7,2,3,5,6,8,9]):

Psi3a:=P2.Psi2d : \

Hb := Feynman\_quantum\_operator(9,"HHH",[1,2,3]):

Psi3b:=Hb.Psi3a :

T4:=Feynman\_quantum\_operator(9,"ccn",[2,3,1]):

Psi31:=CN3.Psi3b :Psi32:=CN2.Psi31 :Psi3:=T4.Psi32 :

{\Large \# Displaying the final state in the Dirac notation \ }

Psi33:=simplify(Feynman\_print(Psi3)):

print(Psi3afterdecoding = Psi33);

end proc:

\bigskip

\end{document}